\documentclass{article}

\usepackage{ifthen}

\usepackage[utf8]{inputenc}
\usepackage[T1]{fontenc}
\usepackage{xspace}
\usepackage{marvosym}
\usepackage{textcomp}
\usepackage[inline]{enumitem}

\usepackage{amsmath}            %
\usepackage{amssymb}			%
\usepackage{amsfonts}			%
\usepackage{nicefrac}			%
\usepackage{amsthm}				%
\usepackage{MnSymbol}			%

\usepackage{mathtools}
\mathtoolsset{mathic=true}			%

\usepackage[linesnumbered,ruled,vlined]{algorithm2e}
\usepackage{comment}
\usepackage{listings}
\usepackage{verbatim}
\usepackage[most]{tcolorbox}

\usepackage{graphicx}
\usepackage{microtype}
\usepackage[
    caption=false,
    font=small,
    ]{subfig}
\usepackage{xcolor}
\usepackage{tikz}
\usetikzlibrary{arrows}
\usetikzlibrary{decorations.pathreplacing}
\usetikzlibrary{spy}
\usepackage{tikz-cd}
\usepackage{tikzit}

\tikzstyle{node}=[fill=black, draw=black, shape=circle, scale=.8]
\tikzstyle{snode}=[fill=black, draw=black, shape=circle, scale=.5]
\tikzstyle{lnodes}=[fill=white, draw=black, shape=circle, very thick, scale=.8]
\tikzstyle{dart}=[fill=white, draw=black, shape=circle, very thick, scale=1.1]
\tikzstyle{ldart}=[fill=none, draw=none, text=black, tikzit fill=white, tikzit draw=black, scale=.8]
\tikzstyle{label}=[fill=white, draw=none, minimum width=4mm, minimum height=3mm, rounded corners=1mm, tikzit fill=white, tikzit draw=black]
\tikzstyle{lgrey}=[fill=none, draw=none, tikzit fill=white, tikzit draw={rgb,255: red,191; green,191; blue,191}, text={rgb,255: red,191; green,191; blue,191}, font={\bfseries}]
\tikzstyle{lrect}=[fill=white, draw=black, very thick, minimum width=6mm, minimum height=4mm, rounded corners=2mm, scale=.8]
\tikzstyle{dtype}=[fill=white, draw=black, very thick, minimum width=6mm, minimum height=4mm, rounded corners=1mm, scale=.8]
\tikzstyle{mline}=[fill=white, draw=black, very thick, align=center, minimum width=6mm, minimum height=4mm, rounded corners=1mm, scale=.8]
\tikzstyle{empty}=[fill=white, draw=black, shape=circle, very thick, minimum width=3mm, minimum height=3mm]
\tikzstyle{steps}=[fill=white, draw=black, shape=circle, very thick, scale=1]
\tikzstyle{l0}=[fill=none, draw=none, text=black]
\tikzstyle{l1}=[fill=none, draw=none, tikzit fill=red, tikzit draw=red, text=red]
\tikzstyle{l2}=[fill=none, draw=none, tikzit fill=blue, tikzit draw=blue, text=blue]
\tikzstyle{l3}=[fill=none, draw=none, tikzit fill={rgb,255: red,0; green,128; blue,0}, tikzit draw={rgb,255: red,0; green,128; blue,0}, text={rgb,255: red,0; green,128; blue,0}]
\tikzstyle{lorange}=[fill=none, draw=none, tikzit fill={rgb,255: red,251; green,131; blue,76}, tikzit draw={rgb,255: red,251; green,131; blue,76}, text={rgb,255: red,251; green,131; blue,76}]
\tikzstyle{lpurple}=[fill=none, draw=none, tikzit fill={rgb,255: red,131; green,76; blue,251}, tikzit draw={rgb,255: red,131; green,76; blue,251}, text={rgb,255: red,131; green,76; blue,251}]
\tikzstyle{rectE}=[fill=none, draw=none, minimum width=4mm, minimum height=3mm, rounded corners=1mm, scale=.6, tikzit fill=white, tikzit draw=none]
\tikzstyle{rect0}=[fill=white, draw=none, minimum width=4mm, minimum height=3mm, rounded corners=1mm, scale=.6, tikzit draw=none]
\tikzstyle{rect1}=[fill=white, draw=none, minimum width=4mm, minimum height=3mm, rounded corners=1mm, scale=.6, tikzit fill=red, tikzit draw=red, text=red]
\tikzstyle{rect2}=[fill=white, draw=none, minimum width=4mm, minimum height=3mm, rounded corners=1mm, scale=.6, tikzit fill=blue, tikzit draw=blue, text=blue]
\tikzstyle{rect3}=[fill=white, draw=none, minimum width=4mm, minimum height=3mm, rounded corners=1mm, scale=.6, tikzit fill={rgb,255: red,0; green,128; blue,0}, tikzit draw={rgb,255: red,0; green,128; blue,0}, text={rgb,255: red,0; green,128; blue,0}]
\tikzstyle{green}=[fill={rgb,255: red,196; green,251; blue,76}, draw={rgb,255: red,196; green,251; blue,76}, shape=circle, scale=.8]
\tikzstyle{UHgren}=[fill={rgb,255: red,196; green,251; blue,76}, draw={rgb,255: red,196; green,251; blue,76}, shape=semicircle, scale=.5]
\tikzstyle{LHgren}=[fill={rgb,255: red,196; green,251; blue,76}, draw={rgb,255: red,196; green,251; blue,76}, shape=semicircle, scale=.5, rotate=180]
\tikzstyle{lrgreen}=[fill={rgb,255: red,196; green,251; blue,76}, draw={rgb,255: red,196; green,251; blue,76}, very thick, minimum width=6mm, minimum height=4mm, rounded corners=2mm, scale=.8]
\tikzstyle{orange}=[fill={rgb,255: red,251; green,131; blue,76}, draw={rgb,255: red,251; green,131; blue,76}, shape=circle, scale=.8]
\tikzstyle{UHorange}=[fill={rgb,255: red,251; green,131; blue,76}, draw={rgb,255: red,255; green,128; blue,76}, shape=semicircle, scale=.5]
\tikzstyle{LHorange}=[fill={rgb,255: red,251; green,131; blue,76}, draw={rgb,255: red,255; green,128; blue,76}, shape=semicircle, scale=.5, rotate=180]
\tikzstyle{lrorange}=[fill={rgb,255: red,251; green,131; blue,76}, draw={rgb,255: red,251; green,131; blue,76}, very thick, minimum width=6mm, minimum height=4mm, rounded corners=2mm, scale=.8]
\tikzstyle{brown}=[fill={rgb,255: red,167; green,93; blue,71}, draw={rgb,255: red,167; green,93; blue,71}, shape=circle, scale=.8]
\tikzstyle{UHbrown}=[fill={rgb,255: red,167; green,93; blue,71}, draw={rgb,255: red,167; green,93; blue,71}, shape=semicircle, scale=.5]
\tikzstyle{LHbrown}=[fill={rgb,255: red,167; green,93; blue,71}, draw={rgb,255: red,167; green,93; blue,71}, shape=semicircle, scale=.5, rotate=180]
\tikzstyle{blue}=[fill={rgb,255: red,76; green,196; blue,251}, draw={rgb,255: red,76; green,196; blue,251}, shape=circle, scale=.8]
\tikzstyle{UHblue}=[fill={rgb,255: red,76; green,196; blue,251}, draw={rgb,255: red,76; green,196; blue,251}, shape=semicircle, scale=.5]
\tikzstyle{LHblue}=[fill={rgb,255: red,76; green,196; blue,251}, draw={rgb,255: red,76; green,196; blue,251}, shape=semicircle, scale=.5, rotate=180]
\tikzstyle{lrblue}=[fill={rgb,255: red,76; green,196; blue,251}, draw={rgb,255: red,76; green,196; blue,251}, very thick, minimum width=6mm, minimum height=4mm, rounded corners=2mm, scale=.8]
\tikzstyle{pink}=[fill={rgb,255: red,251; green,76; blue,196}, draw={rgb,255: red,251; green,76; blue,196}, shape=circle, scale=.8]
\tikzstyle{UHpink}=[fill={rgb,255: red,251; green,76; blue,196}, draw={rgb,255: red,251; green,76; blue,196}, shape=semicircle, scale=.5]
\tikzstyle{LHpink}=[fill={rgb,255: red,251; green,76; blue,196}, draw={rgb,255: red,251; green,76; blue,196}, shape=semicircle, scale=.5, rotate=180]
\tikzstyle{purple}=[fill={rgb,255: red,131; green,76; blue,251}, draw={rgb,255: red,131; green,76; blue,251}, shape=circle, scale=.8]
\tikzstyle{UHpurple}=[fill={rgb,255: red,131; green,76; blue,251}, draw={rgb,255: red,131; green,76; blue,251}, shape=semicircle, scale=.5]
\tikzstyle{LHpurple}=[fill={rgb,255: red,131; green,76; blue,251}, draw={rgb,255: red,131; green,76; blue,251}, shape=semicircle, scale=.5, rotate=180]
\tikzstyle{yellow}=[fill={rgb,255: red,251; green,219; blue,76}, draw={rgb,255: red,251; green,219; blue,76}, shape=circle, scale=.8]
\tikzstyle{UHyellow}=[fill={rgb,255: red,251; green,219; blue,76}, draw={rgb,255: red,251; green,219; blue,76}, shape=semicircle, scale=.5]
\tikzstyle{LHyellow}=[fill={rgb,255: red,251; green,219; blue,76}, draw={rgb,255: red,251; green,219; blue,76}, shape=semicircle, scale=.5, rotate=180]
\tikzstyle{grey}=[fill={rgb,255: red,191; green,191; blue,191}, draw={rgb,255: red,191; green,191; blue,191}, shape=circle, scale=.8]
\tikzstyle{edartgreen}=[fill={rgb,255: red,196; green,251; blue,76}, draw=black, very thick, shape=circle, scale=.6, minimum width=6mm, minimum height=6mm]
\tikzstyle{edartorange}=[fill={rgb,255: red,251; green,131; blue,76}, draw=black, very thick, shape=circle, scale=.6, minimum width=6mm, minimum height=6mm]
\tikzstyle{edartblue}=[fill={rgb,255: red,76; green,196; blue,251}, draw=black, very thick, shape=circle, scale=.6, minimum width=6mm, minimum height=6mm]
\tikzstyle{edartpurple}=[fill={rgb,255: red,131; green,76; blue,251}, draw=black, very thick, shape=circle, scale=.6, minimum width=6mm, minimum height=6mm]
\tikzstyle{edartyellow}=[fill={rgb,255: red,251; green,219; blue,76}, draw=black, very thick, shape=circle, scale=.6, minimum width=6mm, minimum height=6mm]
\tikzstyle{edartwhite}=[fill=white, draw=black, very thick, shape=circle, scale=.6, minimum width=6mm, minimum height=6mm]

\tikzstyle{edge}=[fill=none, very thick, ->, draw=black]
\tikzstyle{mapsto}=[fill=none, draw=black, ->]
\tikzstyle{link}=[fill=none, draw=black, -]
\tikzstyle{intention}=[->, draw={rgb,255: red,191; green,191; blue,191}, fill=none, dashed, dash pattern=on 3mm off 2mm, very thick]
\tikzstyle{green_arrow}=[fill=none, very thick, ->, draw={rgb,255: red,196; green,251; blue,76}]
\tikzstyle{orange_arrow}=[fill=none, very thick, ->, draw={rgb,255: red,251; green,131; blue,76}]
\tikzstyle{brown_arrow}=[fill=none, very thick, ->, draw={rgb,255: red,167; green,93; blue,71}]
\tikzstyle{blue_arrow}=[fill=none, very thick, ->, draw={rgb,255: red,76; green,196; blue,251}]
\tikzstyle{pink_arrow}=[fill=none, very thick, ->, draw={rgb,255: red,251; green,76; blue,196}]
\tikzstyle{purple_arrow}=[fill=none, very thick, ->, draw={rgb,255: red,131; green,76; blue,251}]
\tikzstyle{yellow_arrow}=[fill=none, very thick, ->, draw={rgb,255: red,251; green,219; blue,76}]
\tikzstyle{grey_arrow}=[fill=none, very thick, ->, draw={rgb,255: red,191; green,191; blue,191}]
\tikzstyle{green_double}=[fill=none, very thick, {|->}, draw={rgb,255: red,196; green,251; blue,76}]
\tikzstyle{orange_double}=[fill=none, very thick, {|->}, draw={rgb,255: red,251; green,131; blue,76}]
\tikzstyle{brown_double}=[fill=none, very thick, {|->}, draw={rgb,255: red,167; green,93; blue,71}]
\tikzstyle{blue_double}=[fill=none, very thick, {|->}, draw={rgb,255: red,76; green,196; blue,251}]
\tikzstyle{pink_double}=[fill=none, very thick, {|->}, draw={rgb,255: red,251; green,76; blue,196}]
\tikzstyle{purple_double}=[fill=none, very thick, {|->}, draw={rgb,255: red,131; green,76; blue,251}]
\tikzstyle{yellow_double}=[fill=none, very thick, {|->}, draw={rgb,255: red,251; green,219; blue,76}]
\tikzstyle{orange_bi}=[fill=none, very thick, <->, draw={rgb,255: red,251; green,131; blue,76}]
\tikzstyle{a0}=[fill=none, -, very thick, draw=black]
\tikzstyle{a1}=[fill=none, -, very thick, draw=red]
\tikzstyle{a2}=[fill=none, -, very thick, draw=blue]
\tikzstyle{a3}=[fill=none, -, very thick, draw={rgb,255: red,0; green,128; blue,0}]
\tikzstyle{b0}=[fill=none, ->, very thick, draw=black]
\tikzstyle{b00}=[fill=none, <->, very thick, draw=black]
\tikzstyle{b1}=[fill=none, ->, very thick, draw=red]
\tikzstyle{b11}=[fill=none, <->, very thick, draw=red]
\tikzstyle{vertex}=[->, draw={rgb,255: red,128; green,0; blue,128}, fill=none, dashed, dash pattern=on 2mm off 1mm, very thick]
\tikzstyle{b2}=[fill=none, ->, very thick, draw=blue]
\tikzstyle{b22}=[fill=none, <->, very thick, draw=blue]
\tikzstyle{b3}=[fill=none, ->, very thick, draw={rgb,255: red,0; green,128; blue,0}]
\tikzstyle{b33}=[fill=none, <->, very thick, draw={rgb,255: red,0; green,128; blue,0}]
\tikzstyle{box}=[-, draw={rgb,255: red,191; green,191; blue,191}, fill=none, dashed, dash pattern=on 2mm off 1 mm, very thick]
\tikzstyle{fill_green}=[-, draw={rgb,255: red,196; green,251; blue,76}, fill={rgb,255: red,196; green,251; blue,76}]
\tikzstyle{fill_yellow}=[-, draw={rgb,255: red,251; green,219; blue,76}, fill={rgb,255: red,251; green,219; blue,76}]
\tikzstyle{display}=[fill=none, ->, draw={rgb,255: red,167; green,93; blue,71}]
\tikzstyle{position}=[->, draw={rgb,255: red,115; green,115; blue,115}, fill=none, dashed, dash pattern=on 1mm off .5mm]
\tikzstyle{col}=[->, draw={rgb,255: red,167; green,93; blue,71}, fill=none, dashed, dash pattern=on .66mm off .33mm]

\tikzstyle{lincl}=[fill=none, draw=black, left hook->]
\tikzstyle{rincl}=[fill=none, draw=black, right hook->]
\tikzstyle{dincl}=[fill=none, draw=black, right hook->]
\tikzstyle{uincl}=[fill=none, draw=black, left hook->]
\tikzstyle{partialmono}=[fill=none, draw=black, right hook-left to]
\tikzstyle{functor}=[fill=none, draw=black, -implies, double equal sign distance] \usepackage{quiver}

\usepackage{multirow}

\usepackage{url}
\usepackage{hyperref}
\hypersetup{pdfborder={0 0 0},
			colorlinks=true,
			allcolors=red,
			}

\usepackage[top=3cm, bottom=2.5cm, left=3cm, right=3cm,
			headheight=15pt]{geometry}

\usepackage{orcidlink}

\usepackage{cleveref}
\Crefname{section}{Sec.}{Secs.}
\crefname{section}{Sec.}{Secs.}
\Crefname{definition}{Def.}{Defs.}
\crefname{definition}{Def.}{Defs.}
\Crefname{proposition}{Prop.}{Props.}
\crefname{proposition}{Prop.}{Props.}
\Crefname{example}{Example}{Examples}
\crefname{example}{Example}{Examples}
\Crefname{figure}{Fig.}{Figs.}
\crefname{figure}{Fig.}{Figs.}
\Crefname{appendix}{App.}{Apps.}
\crefname{appendix}{App.}{Apps.}
\Crefname{equation}{Equation}{Equations}
\crefname{equation}{Equation}{Equations}

\usepackage[
	backend=biber,
    style=numeric,
    sorting=nyt,
    date=year,
    giveninits=true,
    maxbibnames=99,
	]{biblatex}
\graphicspath{{fig/}{./}} 

\newtheorem{definition}{Definition}[section]

\mathchardef\mhyphen="2D 													%
\newcommand{\orb}[1]{\langle #1 \rangle}									%
\newcommand{\from}{{\,\colon\,}\linebreak[0]}       							%
\newcommand{\toby}[1]{\xrightarrow{#1}}   									%
\newcommand{\intset}[2]{#1..#2}												%
\newcommand{\set}[1]{\left\{#1\right\}}                                     %

\makeatletter
\def\bstr{b}
\def\bfstr{bf}
\def\cstr{c}
\def\fstr{f}
\def\sstr{s}

\def\strLst{A,B,C,D,d,E,F,G,H,I,J,K,L,M,N,O,P,Q,R,S,T,U,V,W,X,Y,Z}

\newcommand{\MkB}[1]{\expandafter\def\csname\bstr#1\endcsname{\mathbb{#1}}}
	\@for\i:=\strLst\do{%
		\expandafter\MkB \i     }
		
\newcommand{\MkBF}[1]{\expandafter\def\csname\bfstr#1\endcsname{\mathbf{#1}}}
	\@for\i:=\strLst\do{%
		\expandafter\MkBF \i     }

\newcommand{\MkCal}[1]{\expandafter\def\csname\cstr#1\endcsname{\mathcal{#1}}}
	\@for\i:=\strLst\do{%
		\expandafter\MkCal \i     }
		
\newcommand{\MkFrak}[1]{\expandafter\def\csname\fstr#1\endcsname{\mathfrak{#1}}}
	\@for\i:=\strLst\do{%
		\expandafter\MkFrak \i     }
		
\newcommand{\MkSF}[1]{\expandafter\def\csname\sstr#1\endcsname{\mathsf{#1}}}
	\@for\i:=\strLst\do{%
		\expandafter\MkSF \i     }
\makeatother

\definecolor{myorange}{RGB}{251, 131, 76}
\definecolor{mygreen}{RGB}{196, 251, 76}
\definecolor{mypurple}{RGB}{131, 76, 251}
\definecolor{mypink}{RGB}{251, 76, 196}
\definecolor{mybrown}{RGB}{167, 93, 71}
\definecolor{myyellow}{RGB}{251, 219, 76}
\definecolor{myblue}{RGB}{76, 196, 251}
\definecolor{mygray}{RGB}{115, 115, 115}
\definecolor{mylorange}{RGB}{253, 183, 151}
\definecolor{mylgreen}{RGB}{221, 253, 151}
\definecolor{mylpurple}{RGB}{183, 151, 253}
\definecolor{mylpink}{RGB}{253, 151, 221}
\definecolor{mylbrown}{RGB}{197, 136, 118}
\definecolor{mylyellow}{RGB}{253, 234, 151}
\definecolor{mylblue}{RGB}{151, 221, 253}
\definecolor{alpha0}{RGB}{0, 0, 0}
\definecolor{alpha1}{RGB}{255, 0, 0}
\definecolor{alpha2}{RGB}{0, 0, 255}
\definecolor{alpha3}{RGB}{0, 128, 0}
\definecolor{purpledart}{RGB}{128, 0, 128}
\definecolor{pinkdart}{RGB}{255, 0, 255}
\definecolor{lightgray}{gray}{0.9}

\newcommand{\azero}{%
	\textcolor{alpha0}{%
		\ensuremath{0}\xspace%
	}%
}					
\newcommand{\aone}{%
	\textcolor{alpha1}{%
		\ensuremath{1}%
	}\xspace%
}					
\newcommand{\atwo}{%
	\textcolor{alpha2}{%
		\ensuremath{2}%
	}\xspace%
}

\newcommand{\carc}{\rule[2pt]{7pt}{1.5pt}}
\newcommand{\larc}[1]{%
    \raisebox{-.9pt}{%
        $\, \mathbin{\bullet\mkern-3mu{\overset{#1}-}\mkern-3mu\bullet} \,$}
}
\newcommand{\zeroarc}{%
    \mathbin{\bullet\mkern-2mu{\textcolor{alpha0}{\carc}}\mkern-2mu\bullet}
}
\newcommand{\onearc}{%
	\mathbin{\bullet\mkern-2mu{\textcolor{alpha1}{\carc}}\mkern-2mu\bullet}
}
\newcommand{\twoarc}{%
	\mathbin{\bullet\mkern-2mu{\textcolor{alpha2}{\carc}}\mkern-2mu\bullet}
}

\newcommand{\ijpath}[4]{%
    \raisebox{-.9pt}{%
        $\, \mathbin{%
        \bullet
        \mkern-3mu{\overset{#1}-}\mkern-7mu\bullet
        \mkern-7mu{\overset{#2}-}\mkern-7mu\bullet
        \mkern-7mu{\overset{#3}-}\mkern-7mu\bullet
        \mkern-7mu{\overset{#4}-}\mkern-3mu\bullet
        } \,$}
}

\DeclareMathOperator{\pointTWODIM}{\mathtt{Point2}}							%
\DeclareMathOperator{\pointTHREEDIM}{\mathtt{Point3}}						%
\DeclareMathOperator{\colorRGB}{\mathtt{ColorRGB}}							%
\DeclareMathOperator{\ebd}{\pi}												%
\DeclareMathOperator{\pos}{\mathtt{position}}                               %
\DeclareMathOperator{\col}{\mathtt{color}}                                  %

\newcommand{\jerboagraphscheme}[1]{\mathcal{#1}}							%
\newcommand{\jerboarulescheme}[3]{%
	\jerboagraphscheme{#1}
	\toby{#2}
	\jerboagraphscheme{#3}
}

\newlength{\MaxSizeOfLineNumbers}%
\lstdefinestyle{Jerboa}{
	language=Java,
	basicstyle=\scriptsize,
	frame = trbl,
	framesep = 1mm,
	framerule = 1pt,
	frameround = tttt,
	rulecolor = \color[rgb]{0.9,0.9,1},
    xleftmargin=\MaxSizeOfLineNumbers,
	showstringspaces=false,
	extendedchars=true,
	breaklines=true,
    breakatwhitespace=true,
	showtabs=false,
	showspaces=false,
	identifierstyle=\ttfamily,
	keywordstyle=\color[rgb]{0,0,0.75}\bfseries,
	commentstyle=\color[rgb]{0.133,0.545,0.133},
	stringstyle=\color[rgb]{0.627,0.126,0.941},
	backgroundcolor =\color[rgb]{0.98,0.98,1},
    emph={_position,Point3,@0,@1,@2,@3},
    numbers = left,
    escapechar=|,
}

\newtcbox{\codebox}{
	on line, 
	colback=lightgray,
	colframe=lightgray,
	before upper={\rule[-3pt]{0pt}{10pt}},
	boxrule=1pt,
	boxsep=0pt,
	left=2pt,
	right=2pt,
	top=2pt,
	bottom=2pt,
	highlight math style={enhanced}
}
 
\newcommand{\keywords}[1]{%
\textit{Keywords.} #1
}

\title{Instantiation of Jerboa Rule Schemes, a Set-based Explanation}
\author{%
Romain Pascual\,\orcidlink{0000-0003-1282-1933}\\
Karlsruhe Institute of Technology, Karlsruhe, Germany \\
{\it romain.pascual@kit.edu}
}
\date{}

\begin{document}

\maketitle

\begin{abstract}
    This report presents a set-theoretic framework for the instantiation of rule schemes in the Jerboa platform, a tool for developing domain-specific geometric modelers. Jerboa enables the design of geometric modeling operations as graph transformation rules generalized to rule schemes for genericity over the topological content of the operations. Current approaches to algebraic graph transformations are typically described within a finitary \(\cM\)-adhesive category (where \(\cM\) is a suitable system of monomorphisms), employing compositional double-pushout (DPO) semantics for rewriting. In this report, we propose a lightweight, set-theoretic description that exploits the proximity between presheaf topoi and sets to provide an explanation that does not rely on extensive theoretical background. The proposed method simplifies the formal description of modeling operations to bridge the gap between abstract concepts and their practical application in geometric modeling. The framework offers a complementary perspective to categorical approaches at the foundation of Jerboa.

    \medskip
\end{abstract}

\section{Introduction}

This technical report provides a concise, set-theoretic explanation of the instantiation of Jerboa rule schemes. Jerboa~\cite{belhaouari_jerboa_2014} is a platform for creating domain-specific geometric modelers for applications such as mechanical engineering, CAD/CAM, medical imaging, and entertainment. Geometric modelers support creating and editing geometric shapes through modeling operations tailored to the specific domain. Jerboa enables the conception of dedicated modelers by
\begin{enumerate*}[label={(\arabic*)}]
    \item defining topological and geometric properties of the objects, and
    \item designing a collection of modeling operations to modify them.
\end{enumerate*}
Jerboa includes a rule editor where modeling operations are defined as rules in a Domain-Specific Language (DSL) based on graph transformations~\cite{ehrig_fundamentals_2006,heckel_graph_2020}. A graph transformation applies a rule \(L \to R\) to \(G\), where \(L\) and \(R\) are graphs respectively representing the pattern to match and remove and the pattern to insert. For instance, \(L\) might represent a face and \(R\) its subdivision. The standard approach to graph transformation, known as DPO for double pushout, includes an interface \(I\), describing the parts preserved within the rule -- essentially, the intersection of \(L\) and \(R\). Typically, \(L\) specifies an exact match in \(G\), like a specific triangle or quad. To bypass this over-specification, rules have been generalized to rule schemes that abstract the exact topology by an orbit type that essentially describes the topological cell on which the operation can be applied~\cite{pascual_topological_2022}. Applying a rule scheme to a geometric object encoded by a graph \(G\) then requires an instantiation step, where an explicit rule is derived from the rule scheme.
While~\cite{pascual_topological_2022} offers a categorical framework for rule scheme instantiation, the intrinsic proximity between presheaf topoi and sets naturally allows for a set-theoretic explanation of the construction, which is presented here. This report draws on explanations from~\cite{pascual_inferring_2022} and~\cite{pascual_inference_2022}.

\section{Representing Objects as Generalized Maps}

Topology-based geometric modeling deals with an object's outer and complete internal structure, represented by subdividing the object into topological cells. In \(2\)D, objects contain three kinds of cells. Vertices correspond to \(\azero\)-dimensional cells, edges link vertices and form \(\aone\)-dimensional cells, and faces are \(\atwo\)-dimensional cells bordered by edges. In higher dimensions, \(i\)-dimensional cells are enclosed by \((i-1)\)-dimensional cells. For short, \(i\)-dimensional cells are called \(i\)-cells.

In topology-based geometric modeling, the focus is on the topological relations between these cells, which include \emph{adjacency} and \emph{incidence} relations.
An \(i\)-cell is incident to an \((i+1)\)-cell if it belongs to its boundary. In this case, the \((i+1)\)-cell is also said to be incident to the \(i\)-cell. For instance, an edge is incident to a vertex if the vertex is one of its endpoints, while a face is incident to an edge if the edge lies on its boundary. Two \(i\)-cells are adjacent if they share a common cell in their boundary, i.e., if they are incident to the same \((i+1)\)-cell (or \((i-1)\)-cell). For example, two edges are adjacent if they share a vertex or belong to the same face.

\begin{figure}[t!]
    \subfloat[]{%
        \includegraphics[height=42mm]{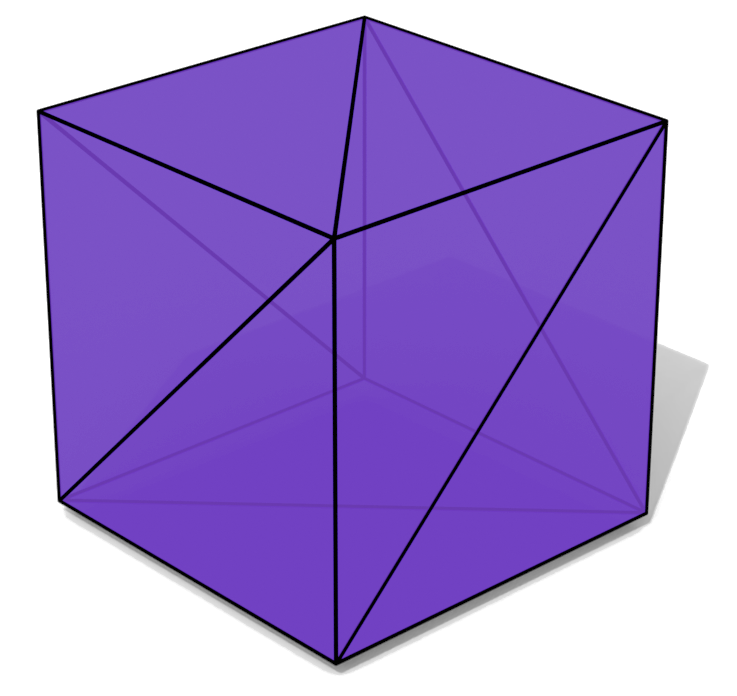}%
        \label{fig:topovsgeom:object}%
 }%
    \hfill%
    \subfloat[]{%
        \includegraphics[height=42mm]{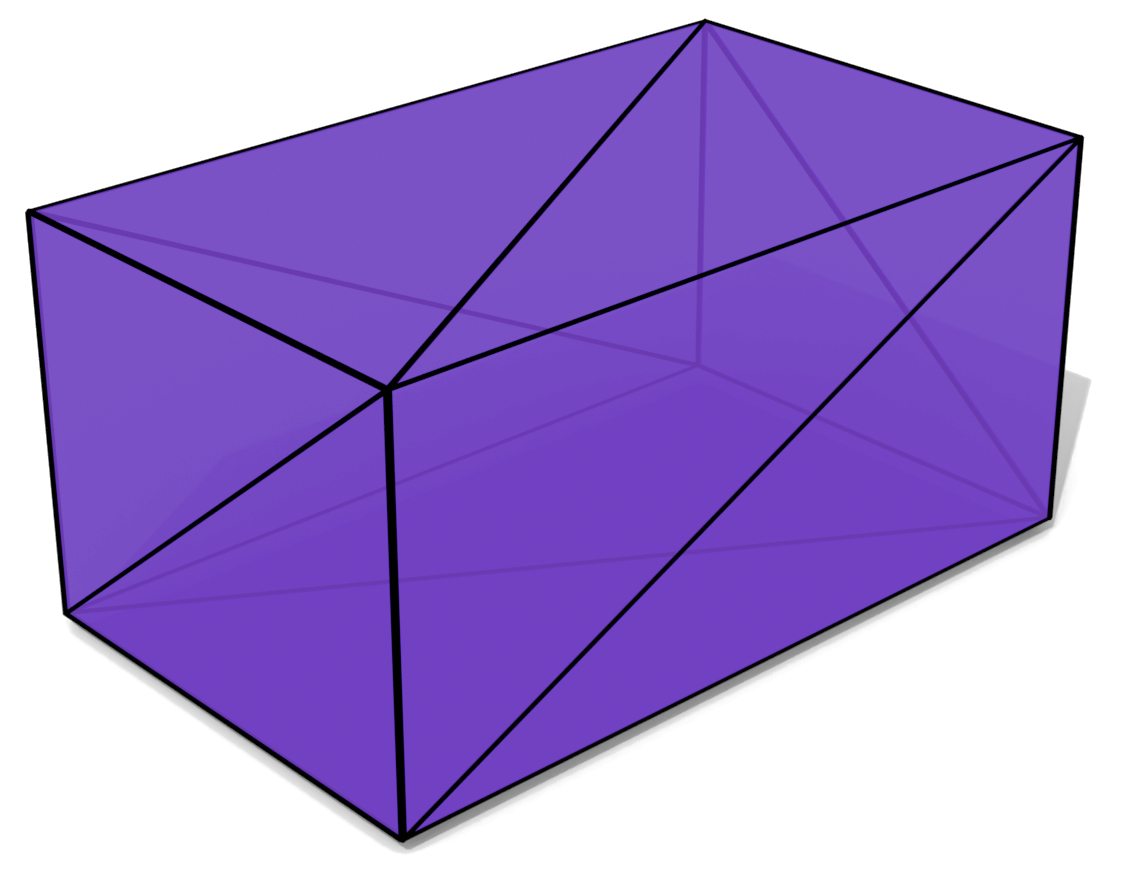}%
        \label{fig:topovsgeom:geom}%
 }%
    \hfill%
    \subfloat[]{%
        \includegraphics[height=42mm]{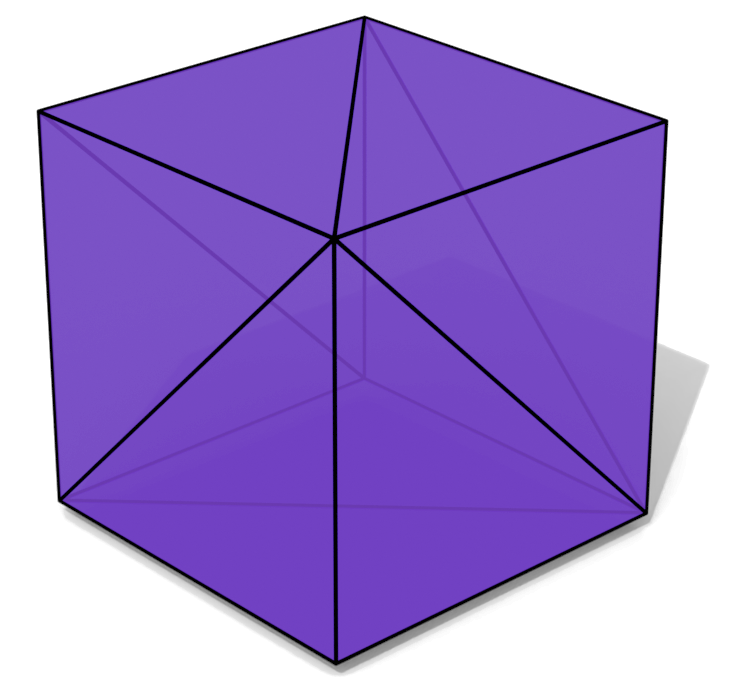}%
        \label{fig:topovsgeom:topo}%
 }%

    \caption[Variations of topology and geometry.]{
 Variations of topology and geometry:
        \subref{fig:topovsgeom:object} an cube,
        \subref{fig:topovsgeom:geom} same topology as \subref{fig:topovsgeom:object} but with a different geometry, and
        \subref{fig:topovsgeom:topo} same geometry as \subref{fig:topovsgeom:object} but with a different topology.}
    \label{fig:topovsgeom}
\end{figure}

The object's topology is completed with geometric information to visualize the object. At the minimum, vertex positions are required to display the object. Additional values such as colors, textures, normals, physical properties, or semantical information may be added based on the application domain. This non-topological information is encapsulated via embedding values added to the topological cells of the object and forms its geometry. The distinction between topology and geometry is illustrated in \cref{fig:topovsgeom}. \Cref{fig:topovsgeom:object} depicts a cube with faces split into two triangles. \Cref{fig:topovsgeom:geom} shows the same topology as \cref{fig:topovsgeom:object} but with a stretched geometry. Conversely, \cref{fig:topovsgeom:topo} maintains the same vertex positions as in \cref{fig:topovsgeom:object} but flips an edge (on the cube's right side), resulting in a different topology.

\subsection{Topological Representation}
\label{subsec:topologicalmodel}

Among the various data structures used to represent geometric objects, edge-based data structures~\cite{weiler_edge-based_1985} encode the exact internal representation of the object. These structures are formally defined as various models of combinatorial maps~\cite{lienhardt_topological_1991}, which define topology through permutations over a set of elements called darts, essentially describing simplicial sets. Among combinatorial maps, generalized maps (Gmaps) allow representing arbitrary cells (e.g., faces with any number of vertices) as explicit substructures, facilitating formal reasoning on subdivided objects. Gmaps have been used for formal verification in Coq~\cite{dehlinger_formal_2014}. A Gmap describes the topology of an object through its cellular decomposition and can be interpreted as a graph with arcs labeled by topological dimensions~\cite{pascual_topological_2022}.

The combinatorial definition of Gmaps~\cite{damiand_combinatorial_2014} uses a set of involutions \(I_1, \ldots, I_n\) over a set of darts \(D\), where each \(I_i\) is a symmetric relation over \(D\), i.e., a subset of \(D \times D\). Thus, the structure \(\langle D, I_1, \ldots, I_n \rangle\) can be interpreted as a graph \((D, I_1 \cup \ldots \cup I_n)\), where each dart corresponds to a node, and each involution \(I_i\) defines a set of \(i\)-labeled undirected arcs. The final set of edges is the union of the \(I_i\)'s, with each arc labeled by its dimension, representing adjacency relations between the object's sub-parts.

In this report, a \emph{graph} is an undirected graph, potentially with parallel arcs and loops. Formally, a graph \(G\) is defined as a triple \((N_G, A_G, \lambda_G)\) where
\(N_G\) is the set of nodes,
\(A_G\) is the set of arcs, and
\(\lambda_G \from A_G \to \powerset_{(1,2)}(N_G)\) is the incidence function.
The incidence function maps each arc \(a\) to a subset of nodes \(\lambda_G(a) \subseteq N_G\) with the condition that \(\lambda_G(a)\) has cardinality \(1\) or \(2\). In the former case, the arc is a loop; in the latter, it connects two nodes.
In the algebraic approach to graph rewriting, the different dimensions are encoded through typing, where a type graph specifies the allowed types of nodes and arcs. For our needs, the typing of the arcs can be seen as a labeling function \(\alpha_G \from A_G \to \Sigma\), where \(\Sigma\) is a labeling alphabet. An arc \(a \in A_G\) with \(\alpha_G(a) = k\) is called a \(k\)-arc, and if nodes \(u\) and \(v\) are connected by an \(k\)-arc, we write \(u \larc{k} v\). The subscript \(_G\) will be omitted.

\begin{definition}[Generalized map -- adapted from~\cite{pascual_topological_2022}]\label{def:gmap}
 A generalized map of dimension \(n\), \(n\)-Gmap, or simply Gmap, is a tuple \((D,L,\lambda,\alpha)\) such that \((D,L,\lambda)\) is a graph and \(\alpha \from L \to \intset{0}{n}\) is a labeling function,\footnote{Where \(\intset{0}{n}\) is the set of integers between \(0\) and \(n\) (included)} satisfying the two following topological constraints:
    \begin{description}
        \item[Incidence constraint] for every dimension \(i\), every dart \(d\) admits a unique incident \(i\)-arc , i.e., 
        \[
            \forall i \in \intset{0}{n}, ~\forall d \in D, ~\exists! l \in L, ~\lambda(l) = d \land \alpha(l) = i.
        \]
        \item[Cycle constraint] for dimensions \(i\) and \(j\) such that \(i+2 \leq j\), any path of length \(4\) labeled by \(ijij\) is a cycle, meaning that the source $u$ and the target $v$ are equal in any path $u \ijpath{i}{j}{i}{j} v$, i.e.,
        \begin{multline*}
            \forall i \in \intset{0}{n}, \forall j \in \intset{0}{n}, ~\forall l_0 \in L, ~\forall l_1 \in L, ~\forall l_2 \in L, ~\forall l_3 \in L,\\
            \left((i+2 \leq j)
            \land \bigwedge_{k = 0,2} (\alpha(l_k) = i)
            \land \bigwedge_{k = 1,3} (\alpha(l_k) = j)
            \land \bigwedge_{k\in\intset{0}{2}} (\lambda(l_k) \cap \lambda(l_{k+1}) \neq \emptyset)\right)
            \Longrightarrow \lambda(l_0) \cap \lambda(l_{3}) \neq \emptyset.
        \end{multline*}
    \end{description}
\end{definition}

\Cref{def:gmap} translates the combinatorial definition of Gmaps~\cite{damiand_combinatorial_2014} into graphs. To avoid vocabulary clashes between communities (graph rewriting, combinatorics, and geometric modeling), we use the combinatorial terms \emph{darts} and \emph{links} for the constitutive elements of a Gmap. The constraints of \cref{def:gmap} are derived from involution properties and ensure that an \((i-1)\)-cell separates at most two \(i\)-cells and that any two \(i\)-cells can only be connected along an \((i-1)\)-cell. For example, in a 2D structure, three faces cannot share an edge, and faces cannot be connected along a vertex. 

\begin{figure}[t!]
    \begin{minipage}{.3\linewidth}
    \subfloat[]{%
        \includegraphics[width=\linewidth]{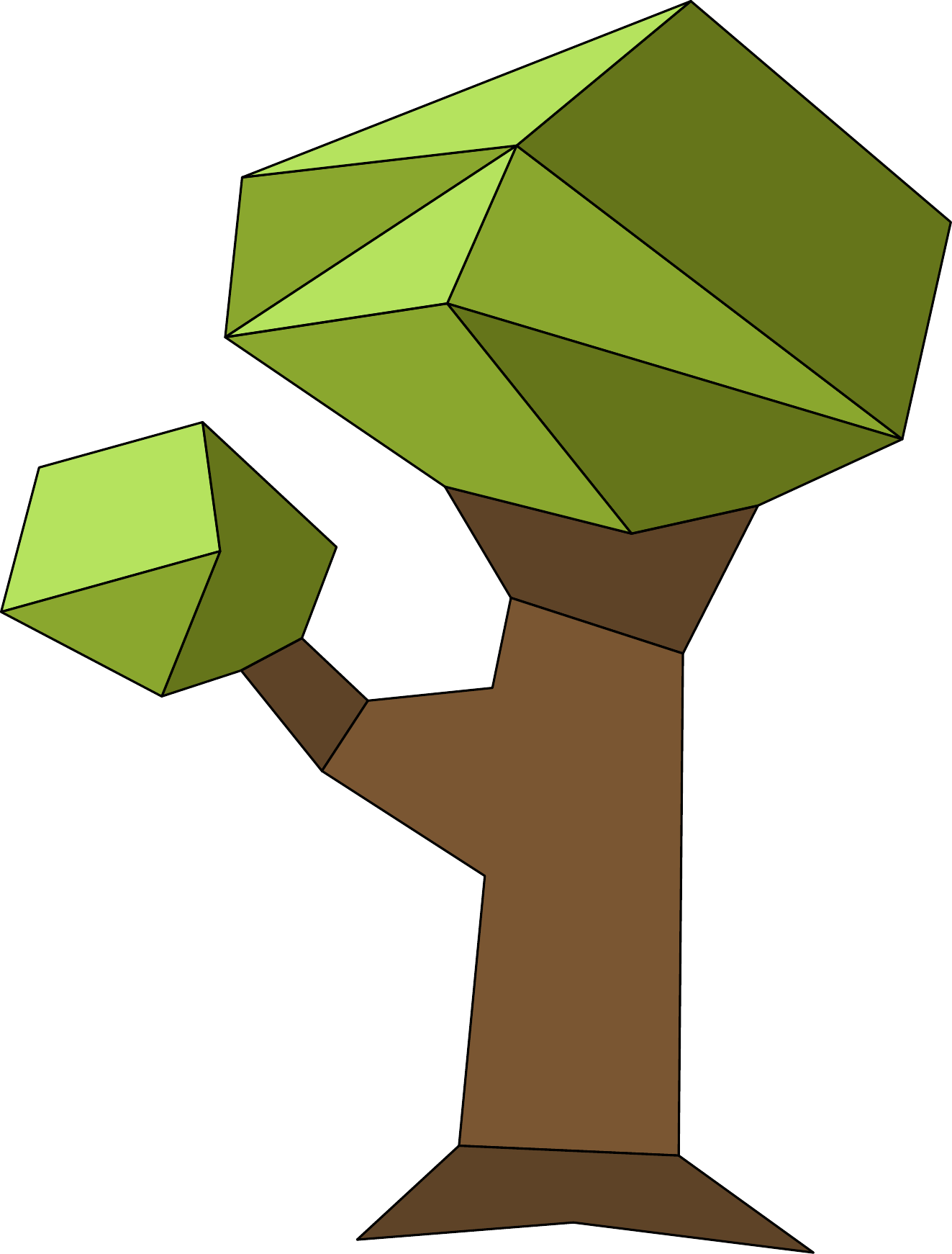}%
        \label{fig:object:1}%
 }%
    \end{minipage}
    \begin{minipage}{.7\linewidth}
    \subfloat[]{%
        \includegraphics[width=.33\linewidth]{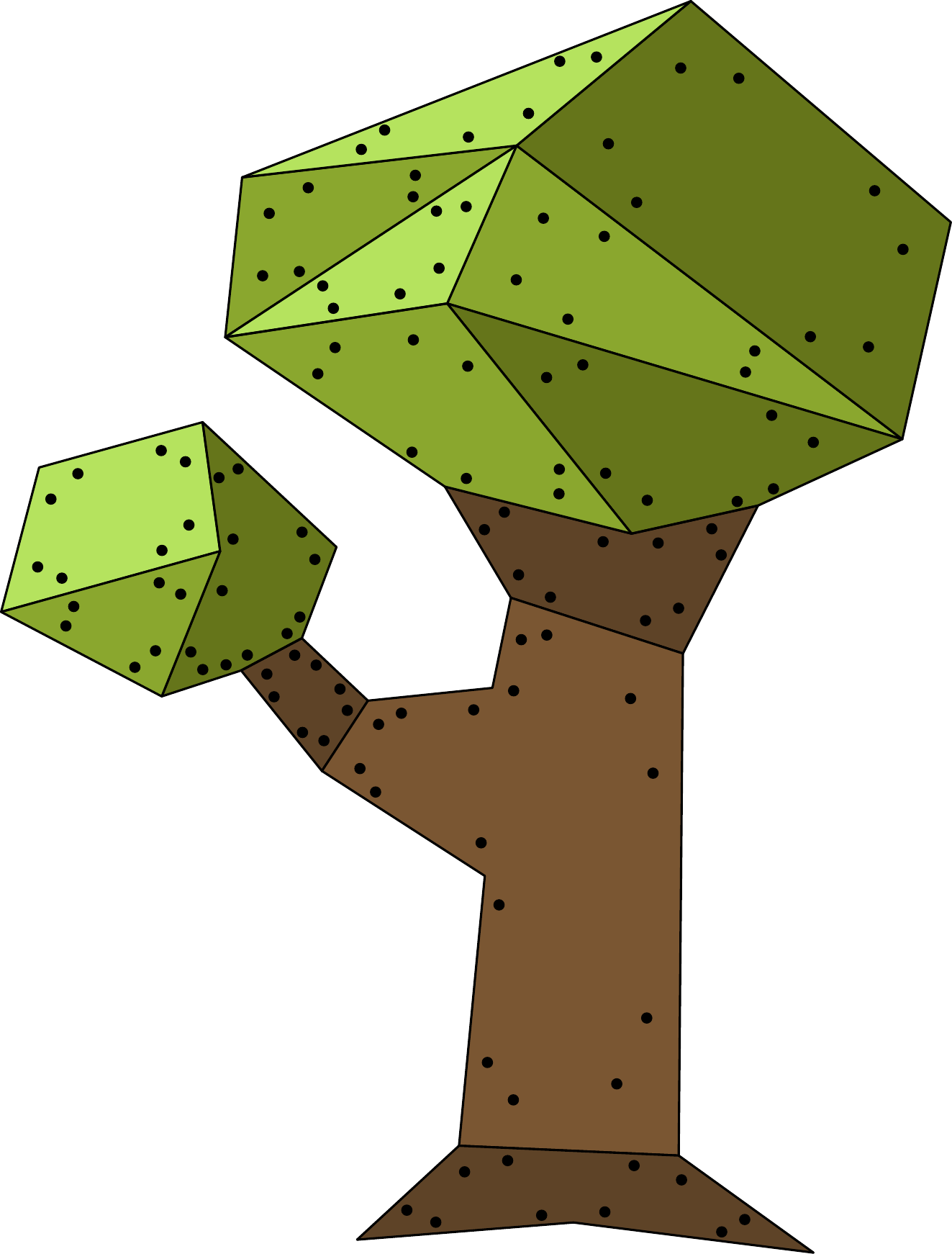}%
        \label{fig:object:2}%
 }%
    \hfill%
    \subfloat[]{%
        \includegraphics[width=.33\linewidth]{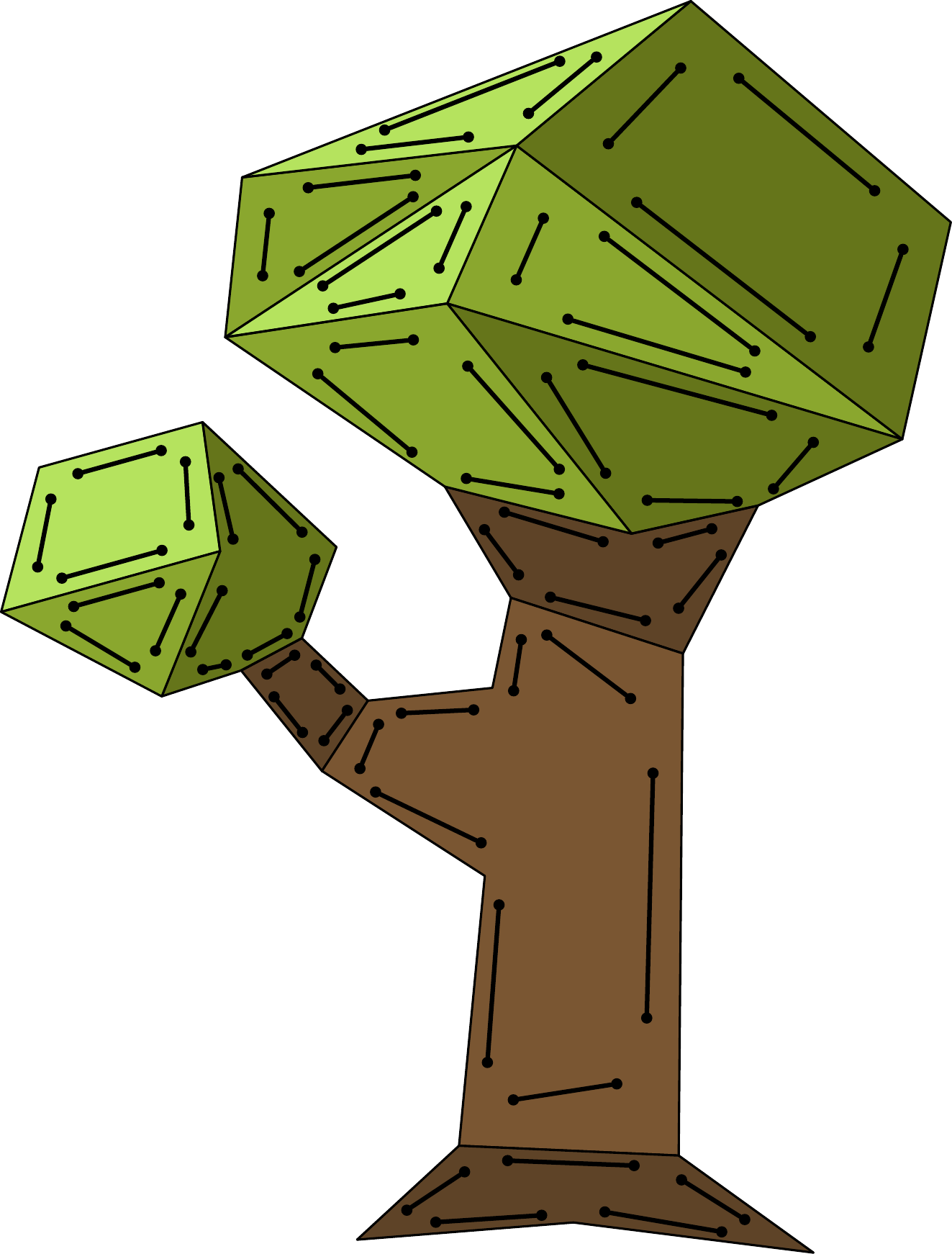}%
        \label{fig:object:3}%
 }%
    \subfloat[]{%
        \includegraphics[width=.33\linewidth]{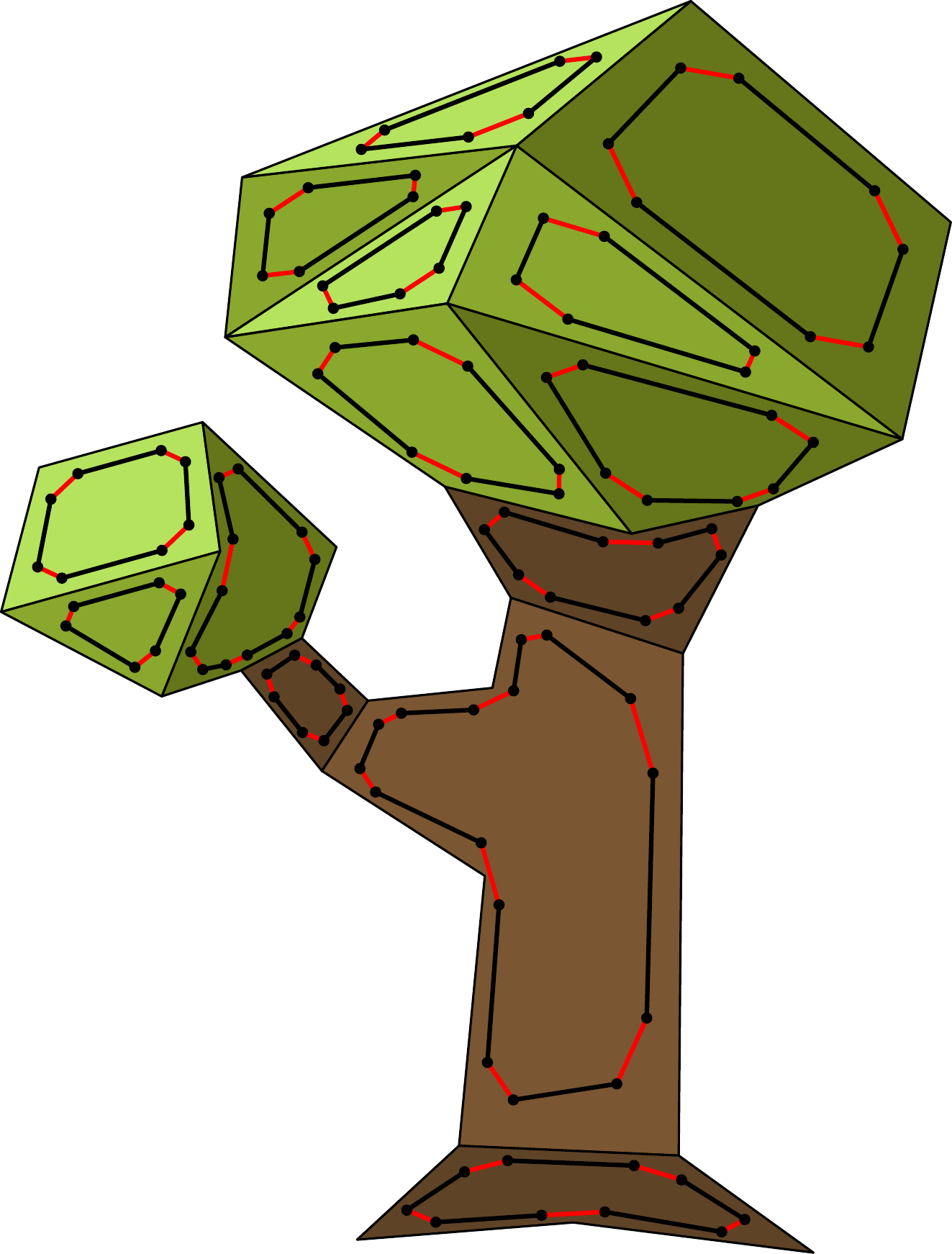}%
        \label{fig:object:4}%
 }%

    \hfill%
    \subfloat[]{%
        \includegraphics[width=.33\linewidth]{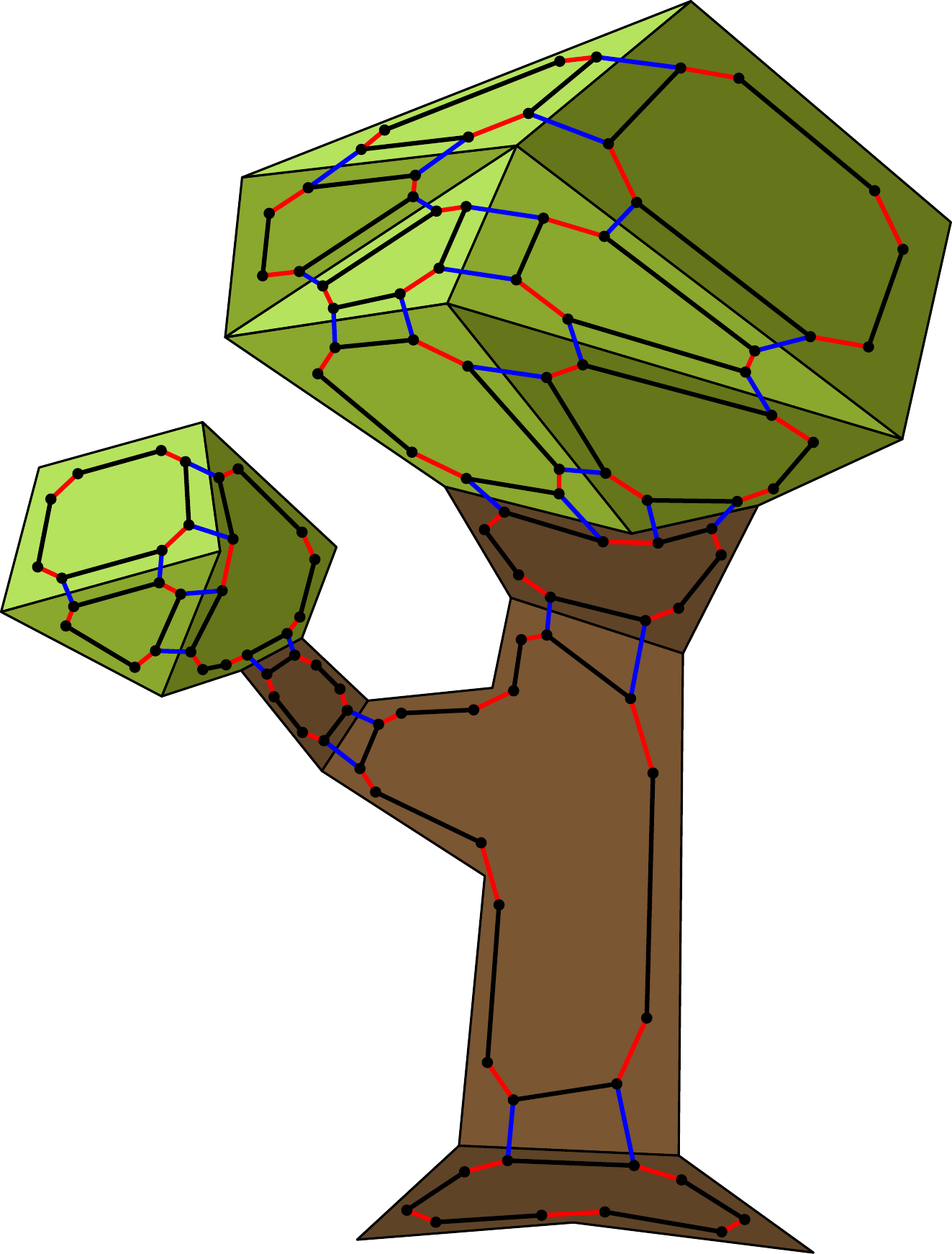}%
        \label{fig:object:5}%
 }%
    \hfill%
    \subfloat[]{%
        \includegraphics[width=.33\linewidth]{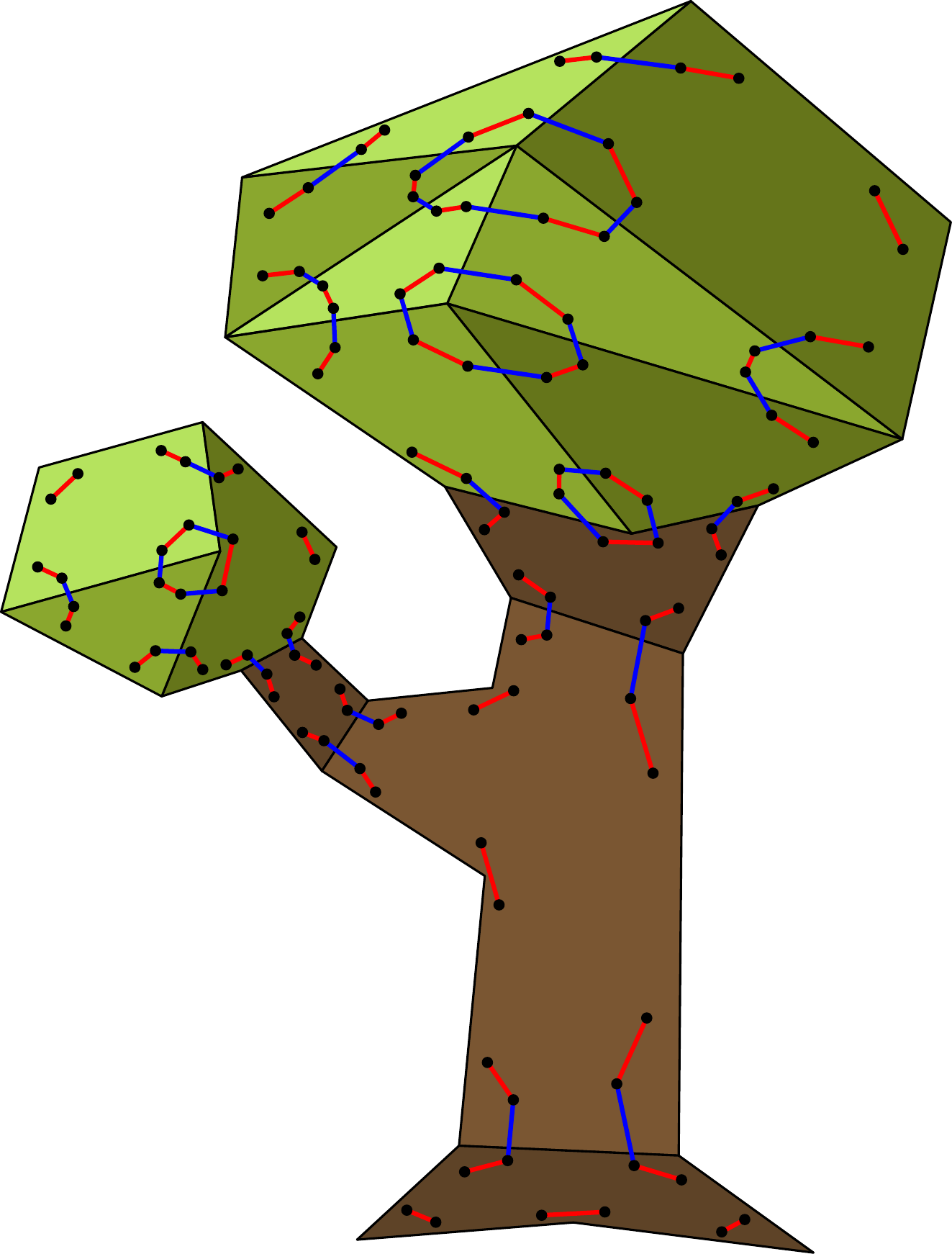}%
        \label{fig:cells:1}%
 }%
    \hfill%
    \subfloat[]{%
        \includegraphics[width=.33\linewidth]{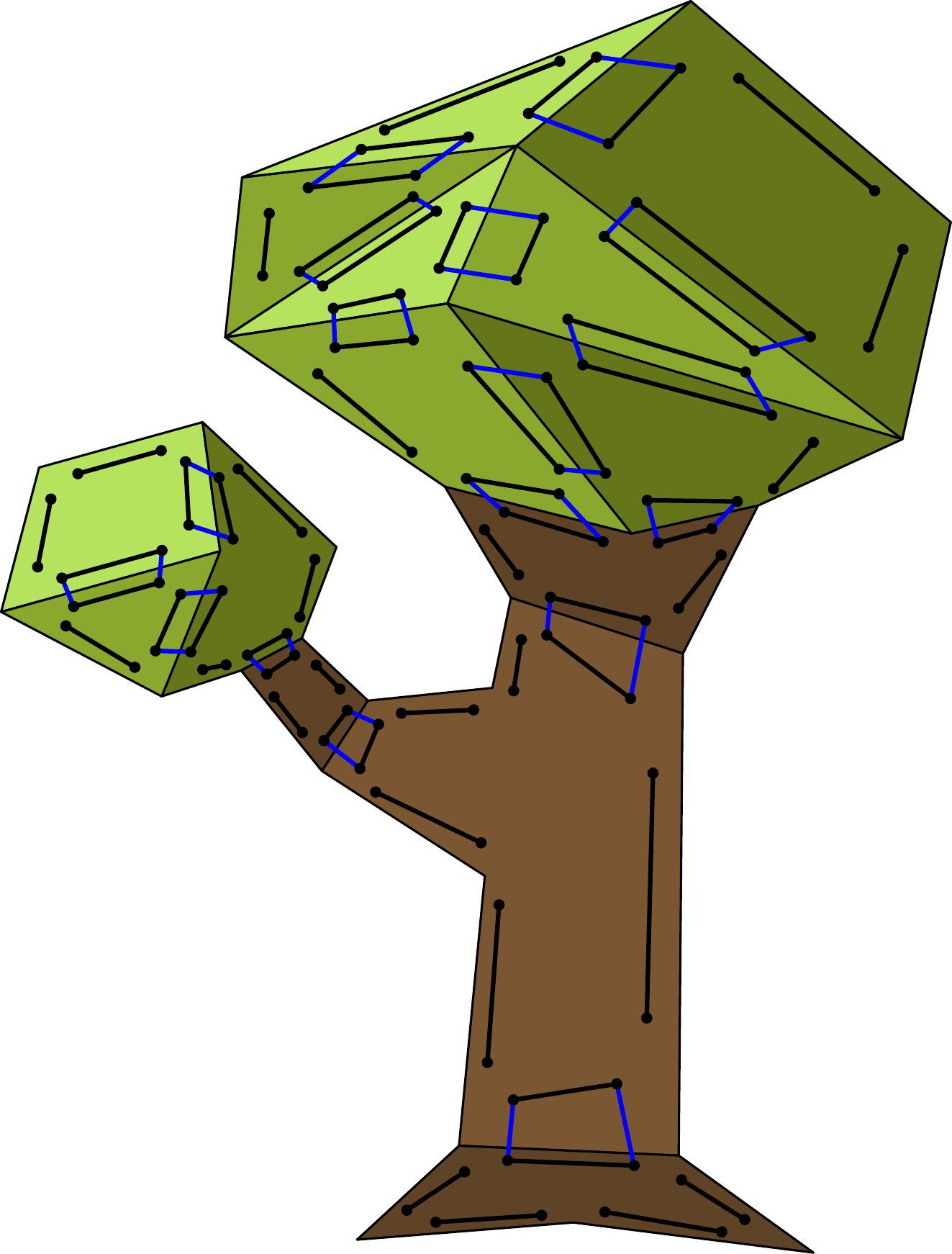}%
        \label{fig:cells:2}%
 }%
    \end{minipage}
    \caption[Topological representation of a geometric object]{%
 Gmap construction:
        \protect\subref{fig:object:1} 2D object,
        \protect\subref{fig:object:2} darts (\(\bullet\)),
        \protect\subref{fig:object:3} \azero-arcs (\(\zeroarc\)),
        \protect\subref{fig:object:4} \aone-arcs (\(\onearc\)),
        \protect\subref{fig:object:5} \atwo-arcs (\(\twoarc\)).
 Cells:
        \protect\subref{fig:cells:1} \(\orb{\aone,\atwo}\)-orbit (vertices),
        \protect\subref{fig:cells:2} \(\orb{\azero,\atwo}\)-orbit (edges),
        \protect\subref{fig:object:4} \(\orb{\azero,\aone}\)-orbit (faces).
 }%
    \label{fig:object}%
\end{figure}

The semantics of the constituent elements of a Gmap allow reconstructing the Gmap corresponding to a given object through a process called \emph{dimensional unification}~\cite{pascual_inference_2022}. Specifically, the semantics of a dart is defined as a tuple of incident cells, and an \(i\)-link encodes the adjacency of two \(i\)-cells sharing a common cell in all other dimensions. For instance, in a \(2\)-Gmap:
\begin{itemize}
    \item A dart is a tuple \((v, e, f)\), where \(v\) is a vertex, \(e\) is an edge, and \(f\) is a face. The tuple satisfies the constraints that \(v\) is an endpoint of \(e\) and \(e\) lies on the boundary of \(f\).
    \item A \(\azero\)-link connects two darts, \(d = (v, e, f)\) and \(d' = (v', e, f)\), which share the same edge \(e\) and face \(f\).
\end{itemize}
An example is illustrated in \cref{fig:object}, showing how a tree is represented as a \(2\)-Gmap through dimensional unification. Beginning with the geometric object in \cref{fig:object:1}, we enumerate all valid dart tuples \((\text{vertex}, \text{edge}, \text{face})\), as shown in \cref{fig:object:2}. These darts are positioned near their corresponding topological cells for ease of visualization. Subsequently, links are incrementally added for each dimension:
\begin{itemize}
    \item \(\azero\)-links (\(\zeroarc\) in black -- \cref{fig:object:3}) connect darts sharing an edge and face but not the same vertex.
    \item \(\aone\)-links (\(\onearc\) in red -- \cref{fig:object:4}) connect darts sharing a vertex and face but not the same edge.
    \item \(\atwo\)-links (\(\twoarc\) in blue -- \cref{fig:object:5}) connect darts sharing a vertex and edge but not the same face.
\end{itemize}

If some dart lacks a sibling for a given \(i\)-link due to their \(i\)-cell lying on the boundary of the object, an \(i\)-loop is added. For simplicity, we consider objects where only the maximal dimension (\(n\) of an \(n\)-Gmap) may have boundaries, resulting in \(n\)-loops. Loops may be omitted from figures for clarity. The graph in \cref{fig:object:5} corresponds to the Gmap derived from the object in \cref{fig:object:1}.

\subsection{Topological Cells and Orbits}
\label{subsec:cell}

We consider an \(n\)-Gmap \(G\). The topological cells (vertices, edges, faces, \ldots) of the associated geometric object are extracted by graph traversal using links of specific dimensions. These subgraphs, parameterized by subsets of dimensions called \emph{orbit types}, are called \emph{orbits}.

\begin{definition}[Orbit -- adapted from~\cite{pascual_topological_2022}]\label{def:orbit}
 An \emph{orbit} of \(G\) is a subgraph formed by all darts reachable from an initial dart through links in a subset \(o \subseteq \intset{0}{n}\). For a dart \(d\), the orbit is denoted as \(G\orb{o}(d)\), or simply \(\orb{o}(d)\) when \(G\) is clear from the context. The orbit is called an \(\orb{o}\)-orbit or said to be of type \(\orb{o}\).
\end{definition}

Intuitively, an orbit groups darts corresponding to a shared topological element. For example, an \(\orb{\aone, \atwo}\)-orbit includes all darts reachable from an initial dart via \(\aone\)- and \(\atwo\)-links. The semantics of darts and links imply that traversing an \(\aone\)-link changes the edge while preserving the vertex and face, whereas traversing an \(\atwo\)-link changes the face while preserving the vertex and edge. Recursively traversing all \(\aone\)- and \(\atwo\)-links from a dart thus allows visiting all darts that belong to the associated vertex, effectively defining the vertex. Orbits describe all topological cells, such as vertices (\(\orb{\azero, \aone}\)-orbits), edges (\(\orb{\azero, \atwo}\)-orbits), and faces (\(\orb{\aone, \atwo}\)-orbits), as shown in \cref{fig:cells:1}, \ref{fig:object:4}, and \ref{fig:cells:2}, respectively.

\subsection{Geometric Information and Embedding}

Gmaps describe the object's topology, describing the decomposition into cells and their connectivity. Although this report emphasizes the topological aspects of modeling operations, we briefly mention how geometric information can be added using embedding functions. These functions assign data such as positions or colors to topological cells, akin to the geometric buffers of a mesh. For instance, vertex positions (on \(\azero\)-cells) or face colors (on \(\atwo\)-cells) provide enough information for the visualization of polyhedral objects.

\begin{figure}[t!]
    \subfloat[]{%
        \includegraphics[width=.32\linewidth]{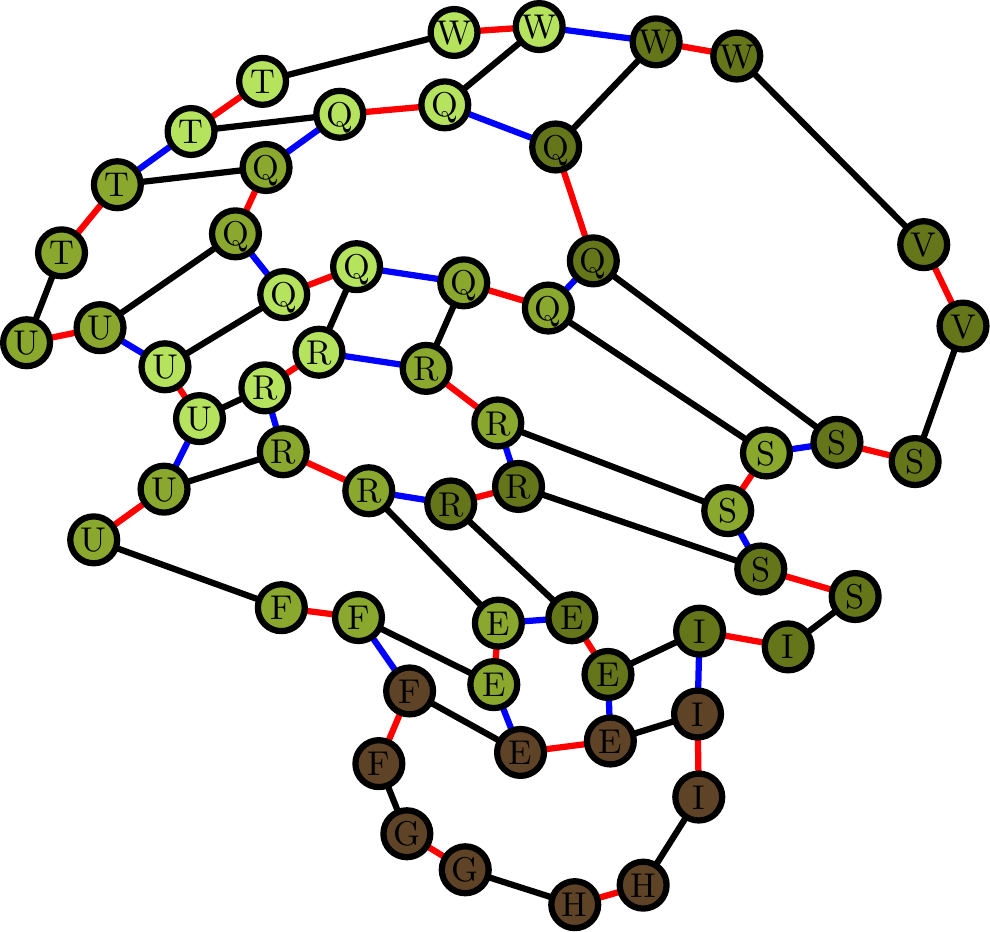}%
        \label{fig:ebd:1}%
 }%
    \hfill%
    \subfloat[]{%
        \includegraphics[width=.32\linewidth]{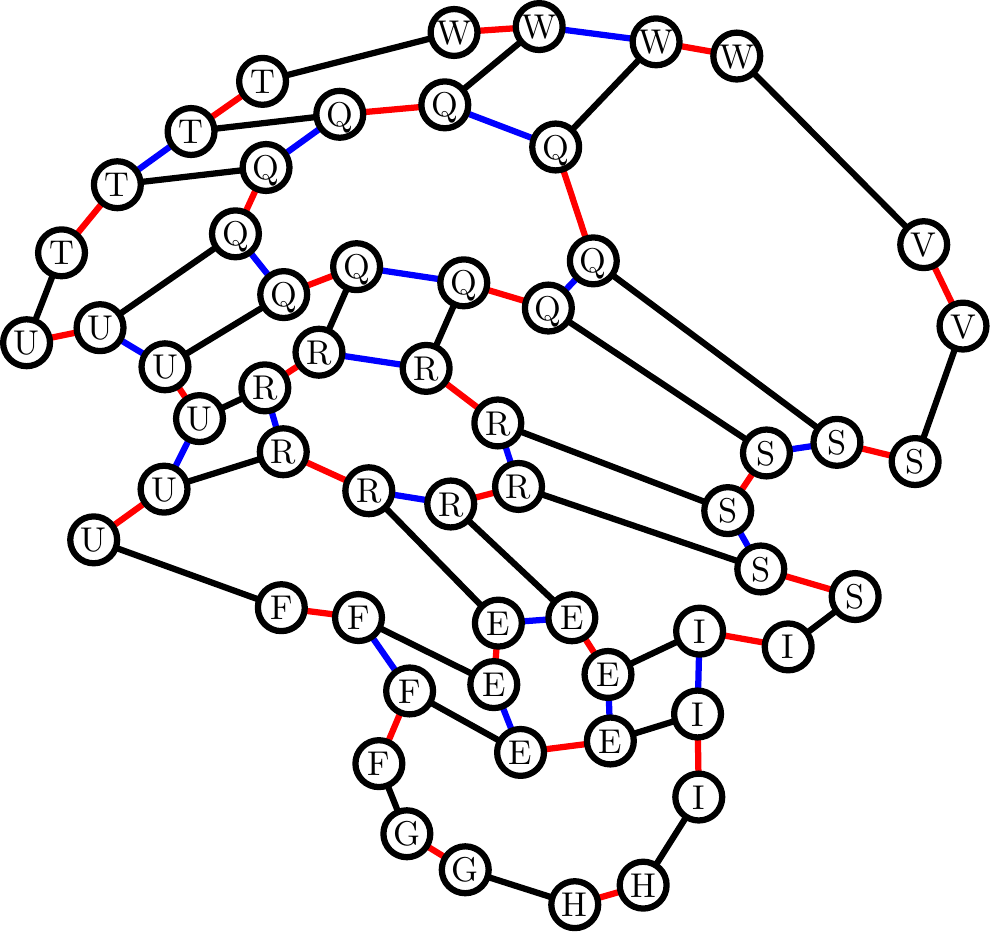}%
        \label{fig:ebd:2}%
 }%
    \hfill%
    \subfloat[]{%
        \includegraphics[width=.32\linewidth]{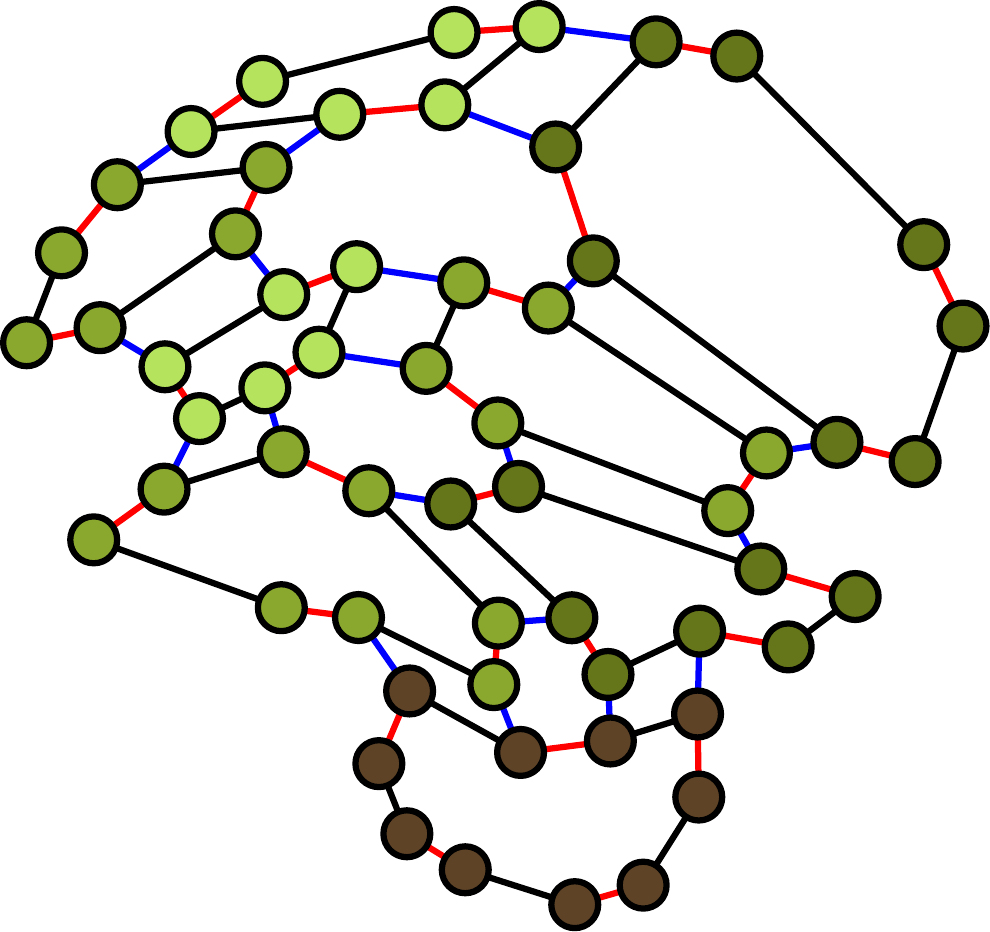}%
        \label{fig:ebd:3}%
 }%
    \caption[Embedding information]{%
    Embeddings:
        \protect\subref{fig:ebd:1} embedded Gmap,
        \protect\subref{fig:ebd:2} \(\pos \from \orb{\aone,\atwo} \to \pointTHREEDIM\),
        \protect\subref{fig:ebd:3} \(\col \from \orb{\azero,\aone} \to \colorRGB\).
 }%
    \label{fig:ebd}%
\end{figure}

Formally, an embedding is a function \(\ebd \from \orb{o_{\ebd}} \to \tau_{\ebd}\), where \(\ebd\) is the embedding name, \(\tau_{\ebd}\) is the data type, and \(\orb{o_{\ebd}}\) is the domain of the embedding~\cite{arnould_preserving_2022}. For example, in a \(2\)-Gmap:
\begin{itemize}
    \item \(\col \from \orb{\azero, \aone} \to \colorRGB\) assigns colors to faces.
    \item \(\pos \from \orb{\aone, \atwo} \to \pointTWODIM\) maps vertices to 2D coordinates.
\end{itemize}
The associated categorical construction is called graph attribution~\cite{ehrig_fundamentals_2006}, but only allows values to be added to nodes or arcs of a graph. 
Embedded Gmaps are obtained by storing a single value for each embedding function on each dart, together with an embedding condition~\cite{arnould_preserving_2022} requiring that all darts within the same \(\orb{o_{\ebd}}\)-orbit share the same \(\tau_{\ebd}\)-value. A portion of the Gmap in \cref{fig:object:1}, representing the tree's top, is shown with embedded positions and colors in \cref{fig:ebd:1}, while the data values are separately illustrated in \cref{fig:ebd:2,fig:ebd:3}.

\section{A Folded Representation of Modeling Operations}
\label{app:instantiation:folded}

From the graph-based definition of Gmaps, modeling operations can be expressed as graph rewriting rules. Rules simplify the design of operations and alleviate their implementation, provided that a suitable rule application engine is available.

\begin{figure}[t!]
        \centering%
        \subfloat[]{%
            \ctikzfig{insertionvertexfree}%
            \label{fig:ajoutSommet1}
 }%
    
        \subfloat[]{%
            \ctikzfig{insertionvertexsewn}%
            \label{fig:ajoutSommet3}
 }%

    \begin{minipage}[b]{.49\linewidth}
        \centering%
        \subfloat[]{%
            \includegraphics[scale=.42]{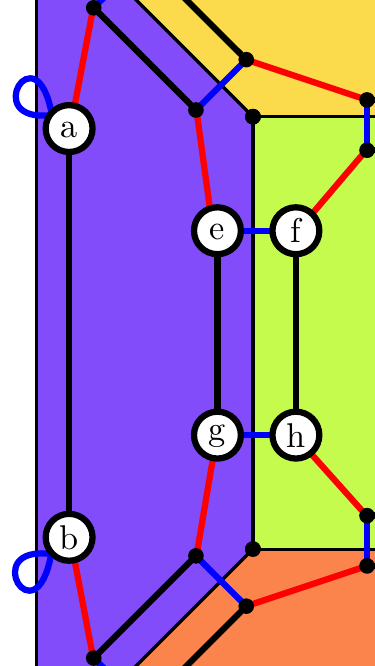}
            \,
            \includegraphics[scale=.42]{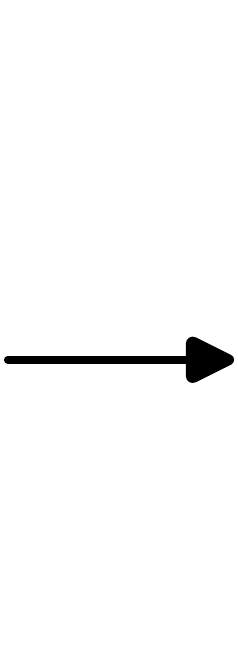}
            \,
            \includegraphics[scale=.42]{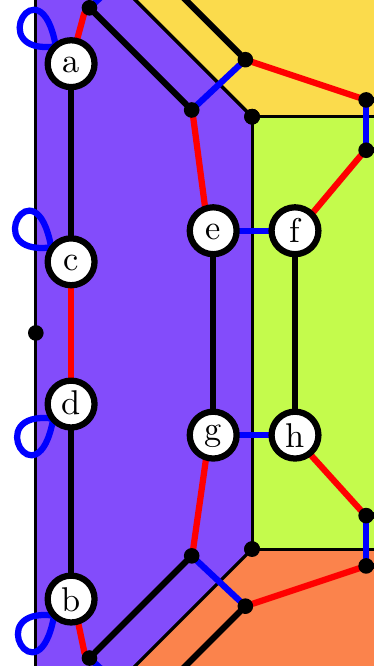}
            \label{fig:ajoutSommet2}
 }%
    \end{minipage}
    \hfill%
    \begin{minipage}[b]{.49\linewidth}
        \centering%
        \subfloat[]{%
            \includegraphics[scale=.42]{vertexinsertion1.pdf}
            \,
            \includegraphics[scale=.42]{vertexinsertionarrow.pdf}
            \,
            \includegraphics[scale=.42]{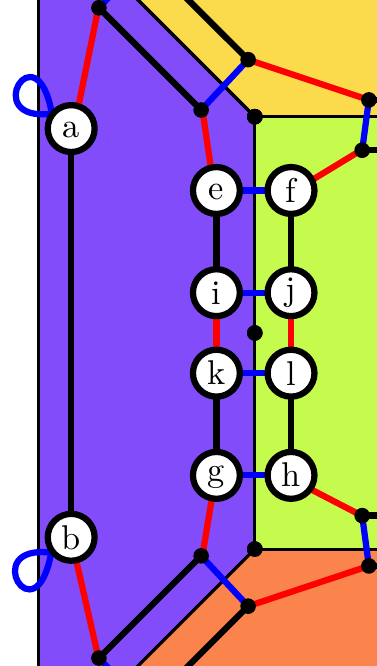}
            \label{fig:ajoutSommet4}
 }%
    \end{minipage}
    
    \caption[Graph transformation rules for the vertex insertion.]{%
 Graph transformation rules for the vertex insertion.
 Graph transformation rule for the vertex insertion in a free edge~\protect\subref{fig:ajoutSommet1} and its application on a \(2\)-Gmap on an outer edge~\protect\subref{fig:ajoutSommet2} via the match deduced from \(x \mapsto a\).
 Graph transformation rule for the vertex insertion in a sewn edge~\protect\subref{fig:ajoutSommet3} and its application on a \(2\)-Gmap on an inner edge~\protect\subref{fig:ajoutSommet4} via the match deduced from \(x \mapsto e\).
 }
    \label{fig:GTongmap}
\end{figure}

Consider the operation of inserting a vertex into an edge. In a \(2\)-Gmap, this operation varies depending on the edge's freedom (i.e., whether it is on the boundary). A free edge forms a \(\orb{\azero, \atwo}\)-orbit with \(\atwo\)-links as loops, while a sewn edge has \(\atwo\)-links as non-loop arcs. This distinction yields two configurations for vertex insertion: one for a free edge (see \cref{fig:ajoutSommet1} for the rule and \cref{fig:ajoutSommet2} for an example application) and one for a sewn edge (see \cref{fig:ajoutSommet3,fig:ajoutSommet4}). These figures zoom in on the modified part, showing both the Gmap and the object for clarity. The object's vertices are marked with dots to highlight the added ones, while darts are annotated with identifiers for readability.

The incidence constraint of Gmaps ensures that specifying a single dart is enough to apply these rules. The complete mapping is then built through a joint traversal of the left-hand side (LHS) and the rewritten graph. 

We illustrate this process using the rule application from \cref{fig:ajoutSommet2}. The match maps node \(x\) to node \(a\). To preserve the node adjacency and the arc labels, the only valid match maps arcs incident to \(x\) onto those incident to \(a\). Thus, \(x \twoarc x\) is mapped to \(a \twoarc a\), and \(x \zeroarc y\) to \(a \zeroarc b\). Mapping these arcs further constrains the match, such that \(y\) must map to \(b\). By recursively exploring incident arcs of each newly mapped node, we can reconstruct the full match starting from the mapping of \(x\) to \(a\). The incidence constraint allows deriving the complete match from an initial mapping of one node per connected component of \(L\).

\begin{figure}[t!]
    \centering%
    \subfloat[]{%
        \includegraphics[scale=.7]{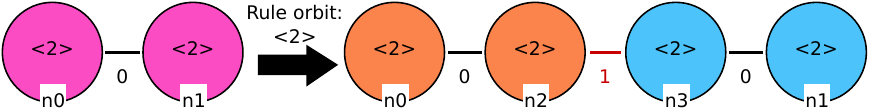}
        \label{fig:vertexinsertion2folded}
 }%
    
    \subfloat[]{%
        \includegraphics[scale=.7]{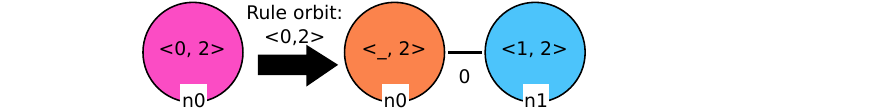}
        \label{fig:vertexinsertion02folded}
 }%

    \caption[rule schemes for the vertex insertion.]{%
 rule schemes for the vertex insertion:
        \protect\subref{fig:vertexinsertion2folded} by folding the \atwo-links and
        \protect\subref{fig:vertexinsertion02folded} both the \azero- and \atwo-links.
 }
    \label{fig:vertexinsertion}
\end{figure}

A rule scheme provides a compact, folded representation of a transformation, which is unfolded to obtain the specific graph transformation applicable to an object. For example, the two configurations for vertex insertion can be unified by folding the edge along its \atwo-links, yielding the rule scheme shown in \cref{fig:vertexinsertion2folded} and parameterized by the orbit type \(\orb{\atwo}\). To create a graph-level rule, we start with a graph consisting of an \(\orb{\atwo}\)-orbit, using it to unfold node decorations. If unfolded as a \atwo-loop, the rule scheme yields the rule in \cref{fig:ajoutSommet1}; unfolding it with two darts sharing a \atwo-link gives the rule in \cref{fig:ajoutSommet3}. Further folding produces the rule scheme shown in \cref{fig:vertexinsertion02folded}. These folding and unfolding processes are defined using relabeling functions.

\section{Relabeling Function}
\label{app:instantiation:relabeling}

\begin{definition}[Relabeling function -- from~\cite{pascual_inferring_2022}]
 A relabeling function of dimension \(n\), or relabeling function, is a partial function \(f\from \intset{0}{n} \to \intset{0}{n} \cup \{ \_ \}\), injective on \(\intset{0}{n}\). Here, `\(\_\)' is a special symbol called the removing symbol.
\end{definition}

Applying a relabeling function to an orbit type involves applying it to each dimension within the orbit. For example, \(\{\azero \mapsto \aone, \atwo \mapsto \atwo\}(\orb{\azero, \atwo}) = \orb{\aone, \atwo}\). The positions of dimensions in the orbit type fully describe the relabeling function, which can then be recovered from reference orbit type \(\orb{o}\) and its relabeled version \(\orb{o'}\). For instance, given \(\orb{\azero, \atwo} \mapsto \orb{\aone, \atwo}\), the relabeling function \(\{\azero \mapsto \aone, \atwo \mapsto \atwo\}\) can be reconstructed unambiguously. More precisely, the motivation behind relabeling functions is to encode orbit rewriting, so we often denote them as relabelings of orbit types. Let \(\orb{o} = (o_i)_{i \leq k}\) be the set of dimensions on which \(f\) is defined (ordered by increasing values); then \(f\) is expressed as \(\orb{o} \mapsto \orb{(f(o_i))_{i \leq k}}\). The injectivity constraint ensures that no dimension \(d\) appears more than once in \(\orb{o'} = \orb{(f(o_i))_{i \leq k}}\). Note that \(\orb{o'}\) is not strictly an orbit type, as it may contain the symbol `\(\_\)' and thus is referred to as a generalized orbit type, which we will call simply an orbit type for convenience. The relabeling function's domain \(\orb{o}\) must not include the removing symbol. A relabeling function naturally extends from orbit type rewriting to orbit rewriting. Given a relabeling function \(\orb{o} \mapsto \orb{o'}\) and an orbit \(\orb{o}(v)\), the orbit \(\orb{o'}(v)\) is obtained by relabeling all links according to the function.

\begin{figure}[t!]
    \centering%
    \subfloat[]{\label{fig:orbit1free}%
    \includegraphics[scale=.3]{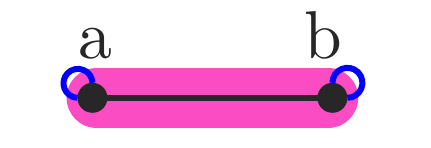}}%
    \hspace{1ex}%
    \subfloat[]{\label{fig:orbitrelabelingfree}%
    \includegraphics[scale=.3]{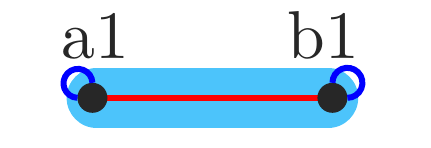}}%
    \hspace{1ex}%
    \subfloat[]{\label{fig:orbitdeletionfree}%
    \includegraphics[scale=.3]{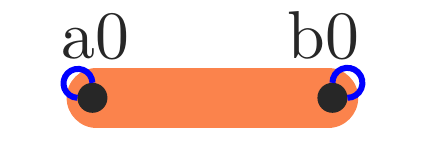}}%
    
    \subfloat[]{\label{fig:orbit1sewn}%
    \includegraphics[scale=.3]{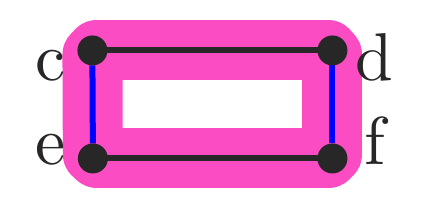}}%
    \hspace{1ex}%
    \subfloat[]{\label{fig:orbitrelabelingsewn}%
    \includegraphics[scale=.3]{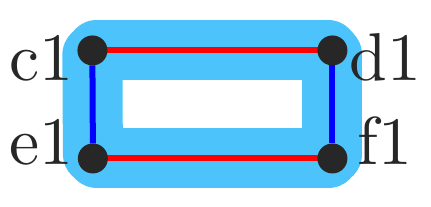}}%
    \hspace{1ex}%
    \subfloat[]{\label{fig:orbitdeletionsewn}%
    \includegraphics[scale=.3]{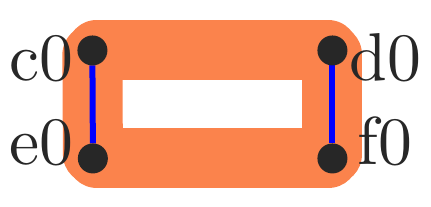}}%
     
    \caption[Relabeling functions applied to orbit graphs.]{%
 Relabeling functions applied to orbit graphs:
 orbits~\protect\subref{fig:orbit1free} and~\protect\subref{fig:orbit1sewn} of type \(\orb{\azero, \atwo}\), label modification~\protect\subref{fig:orbitrelabelingfree} and~\protect\subref{fig:orbitrelabelingsewn} via the relabeling function \(\orb{\azero, \atwo} \mapsto \orb{\aone, \atwo}\), and label deletion~\protect\subref{fig:orbitdeletionfree} and~\protect\subref{fig:orbitdeletionsewn} via the relabeling function \(\orb{\azero, \atwo} \mapsto \orb{\_, \atwo}\).}
    \label{fig:relabeling}
\end{figure}

\Cref{fig:relabeling} illustrates the application of relabeling functions to orbit graphs.
Applying the relabeling function \(\{\azero \mapsto \aone, \atwo \mapsto \atwo\}\) to the graph in \cref{fig:orbit1free} yields the graph in \cref{fig:orbitrelabelingfree}.
The highlighted parts in \cref{fig:relabeling} will be exploited later; for now, we focus on the relabeling of links, i.e., the modifications of link colors.
The \atwo-loops incident to nodes \(a\) and \(b\) become \atwo-loops incident to nodes \(a1\) and \(b1\), via the relabeling \(\atwo \mapsto \atwo\). Similarly, the relabeling \(\azero \mapsto \aone\) transforms \(a \zeroarc b\) into \(a1 \onearc b1\). 
Applying this relabeling function to the graphs in \cref{fig:orbit1sewn} yields the graphs in \cref{fig:orbitrelabelingsewn}.

As the name suggests, the removing symbol `\(\_\)' signifies the deletion of the relabeled dimension, extending the definition of relabeling functions and their application to orbits. For instance, the relabeling function \(\{\azero \mapsto \_, \atwo \mapsto \atwo\}\) indicates the removal of \azero{} while preserving \atwo. The removing symbol may appear in node decorations, such as \(\orb{\_, \atwo}\). Given the reference orbit type \(\orb{\azero, \atwo}\), one can reconstruct the relabeling function \(\{\azero \mapsto \_, \atwo \mapsto \atwo \}\) without ambiguity. When applied to an orbit, links relabeled with `\(\_\)' are deleted. 
Figures \cref{fig:orbitdeletionfree} and \cref{fig:orbitdeletionsewn} illustrate the deletion with the graphs from \cref{fig:orbit1free} and~\ref{fig:orbit1sewn}. In these examples, the links \(a \zeroarc b\), \(c \zeroarc d\), and \(e \zeroarc f\) are removed.

Formally, given a labeled graph \(H = (D,L,\lambda,\alpha)\), and a relabeling function \(f\), the application of the relabeling function \(f\) to \(H\), written \(f(H)\), yields the graph \((D,A',\lambda',\alpha')\) such that
\begin{itemize}
    \item \(L' = \set{l \in L \mid f \circ \alpha(l) \neq \_}\),
    \item \(\lambda'\) is the restriction of \(\lambda\) to \(L'\),
    \item \(\alpha' \from L' \to \intset{0}{n}\) is the function \(l \mapsto f \circ \alpha(l)\) for any \(l\) in \(L'\).
\end{itemize}

Relabeling functions allow encoding folded representation of graphs and rules as graph and rule schemes.

\begin{definition}[Graph scheme, rule scheme -- from~\cite{pascual_inferring_2022}]
    \label{def:jerboaschemes}
 Let \(\orb{o}\) be an orbit type on \(\intset{0}{n}\).

 A graph scheme of dimension \(n\) on \(\orb{o}\), \((n,\orb{o})\)-graph scheme, or simply graph scheme, is a graph \(\jerboagraphscheme{G}\) whose arcs are labeled on \(\intset{0}{n}\) and nodes are decorated with generalized orbit types of the same size as \(\orb{o}\).
 For each node \(\mu\) in \(\jerboagraphscheme{G}\), we denote the orbit type decorating \(\mu\) by \(\orb{o^\mu}\).

 A rule scheme on \(\orb{o}\), or simply rule scheme, is a rule \(\jerboarulescheme{L}{\orb{o}}{R}\) where \(\jerboagraphscheme{L}\) and \(\jerboagraphscheme{R}\) are graph schemes on \(\orb{o}\).
\end{definition}

The orbit type of a rule scheme is also referred to as its parameter and may be omitted when the context is clear. Both graph schemes in a rule scheme must share the same orbit type. The node decorations of a graph scheme are placeholders to represent any orbit of the given orbit type. More specifically, each node in a graph scheme is intended to be substituted by an orbit whose type matches its decoration.
A graph scheme reduced to a single node decorated with the orbit type \(\orb{\azero, \atwo}\) describes both a free or sewn edge, as shown in \cref{fig:orbit1free} and~\ref{fig:orbit1sewn}. Other examples of graph schemes are found in \cref{fig:discretegraphscheme,ex:instantiationarc:fig:graphscheme}.
An additional condition applies when the graph scheme \(\jerboagraphscheme{G}\) contains multiple nodes. The size condition on the orbit types decorating the nodes of \(\jerboagraphscheme{G}\) requires that they all share the same number of symbols as \(\orb{o}\). Thus, all nodes in \(\jerboagraphscheme{G}\) can be substituted based on the same orbit type \(\orb{o}\), with their links relabeled according to the relabeling function derived from the node decorations. Examples are given in \cref{fig:vertexinsertion}.

\section{Instantiation}
\label{app:instantiation:instantiation}

We now explain the instantiation process using the previously introduced relabeling functions. Unfolding a graph scheme through relabeling functions, i.e., instantiating it, is defined separately for nodes and arcs.
In this section, we consider a graph scheme \(\jerboagraphscheme{G}\) on the orbit type \(\orb{o}\) and a graph \(O\) corresponding to an orbit typed by \(\orb{o}\), i.e., \(O = (D_O, L_O, \lambda_O, \alpha_O)\) such that \(O = O\orb{o}(d)\) for any dart \(d\) in \(D_O\). We detail how to obtain the instantiation of \(\jerboagraphscheme{G}\) with the orbit graph \(O\), written \(\iota^{\orb{o}}(\jerboagraphscheme{G}, O)\).

\subsection{Nodes}
The first component is the instantiation of a node \(s\) of \(\jerboagraphscheme{G}\), given by applying the relabeling function \(\orb{o} \mapsto \orb{o^s}\) to the orbit graph \(O\). All nodes can then be instantiated by applying their respective relabeling functions to copies of O.

\begin{figure}[t!]
    \begin{minipage}[b]{0.3\linewidth}
        \centering%
        \subfloat[]{\label{fig:discretegraphscheme}%
        \includegraphics[scale=.7]{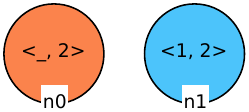}}%
    \end{minipage}
    \begin{minipage}[b]{0.3\linewidth}
        \centering%
        \subfloat[]{\label{fig:discretegraphinstantiationfree}%
        \includegraphics[scale=.3]{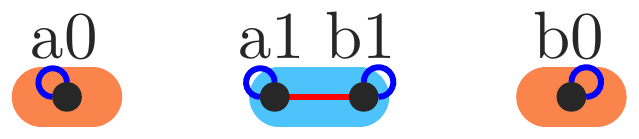}}%
    \end{minipage}
    \begin{minipage}[b]{0.3\linewidth}
        \centering%
        \subfloat[]{\label{fig:discretegraphinstantiationsewn}%
        \includegraphics[scale=.3]{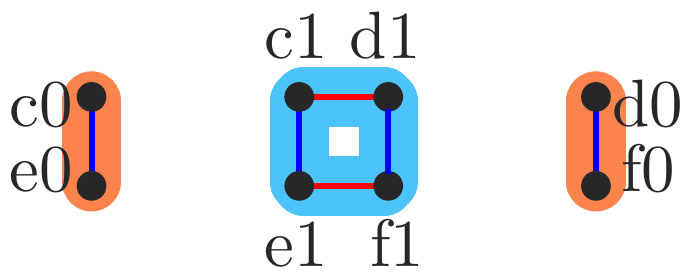}}%
    \end{minipage}

    \caption[Instantiating the nodes of a graph scheme.]{%
 Instantiating the nodes of a graph scheme:
    \protect\subref{fig:discretegraphscheme} discrete graph scheme,
    \protect\subref{fig:discretegraphinstantiationfree} instantiation with the orbit graph of \cref{fig:orbit1free}, and
    \protect\subref{fig:discretegraphinstantiationsewn} instantiation with the orbit graph of \cref{fig:orbit1sewn}.}
    \label{fig:instantiationDiscrete}
\end{figure}

\Cref{fig:discretegraphscheme} shows the two nodes of the RHS of the rule scheme from \cref{fig:vertexinsertion02folded}.
We consider their instantiation on the orbit type \(\orb{o}=\orb{\azero,\atwo}\).

Node \(n0\) has the orbit type \(\orb{\_,\atwo}\), meaning that the corresponding relabeling function is \(\orb{o} \mapsto \orb{o^{n0}} = \{\azero \mapsto \_, \atwo \mapsto \atwo\}\). Similarly, node \(n1\) has the orbit type \(\orb{\aone, \atwo}\), yielding the relabeling function \(\orb{o} \mapsto \orb{o^{n1}} = \{\azero \mapsto \aone, \atwo \mapsto \atwo\}\). These functions, already shown in \cref{fig:relabeling}, allow the instantiations for a free or sewn edge, which are given in \cref{fig:discretegraphinstantiationfree,fig:discretegraphinstantiationsewn}.

\begin{definition}[Node instantiation -- from~\cite{pascual_inferring_2022}]
    \label{def:nodeinstantiation}
 If \(\mu\) is a node of \(\jerboagraphscheme{G}\), its instantiation with \(O\) is the graph obtained by applying \(\orb{o} \mapsto \orb{o^\mu}\) to \(O\):
    \[
        \iota^{\orb{o}}(\mu, O) = [\orb{o} \mapsto \orb{o^\mu}](O).
    \]

 The construction extends to the set of nodes \(N_{\jerboagraphscheme{G}}\) of \(\jerboagraphscheme{G}\), whose instantiation is the union of the instantiations of each node:
    \[
        \iota^{\orb{o}}(N_{\jerboagraphscheme{G}}, O) = \bigcup_{\mu \in N_{\jerboagraphscheme{G}}} \iota^{\orb{o}}(\mu, O).
    \]
\end{definition}

\subsection{Arcs}
The instantiation of an arc between two nodes adds links between darts that are images of the same initial dart via the two relabeling functions.

\begin{figure}[t!]
    \begin{minipage}[b]{0.3\linewidth}
        \centering%
        \subfloat[]{\label{ex:instantiationarc:fig:graphscheme}%
        \includegraphics[scale=.7]{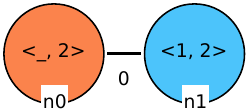}}%
    \end{minipage}
    \begin{minipage}[b]{0.3\linewidth}
        \centering
        \subfloat[]{\label{ex:instantiationarc:fig:instantiationfree}%
        \includegraphics[scale=.3]{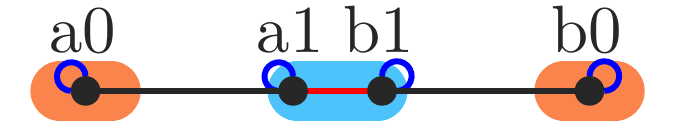}}%
    \end{minipage}
    \begin{minipage}[b]{0.3\linewidth}
        \centering%
        \subfloat[]{\label{ex:instantiationarc:fig:instantiationsewn}
        \includegraphics[scale=.3]{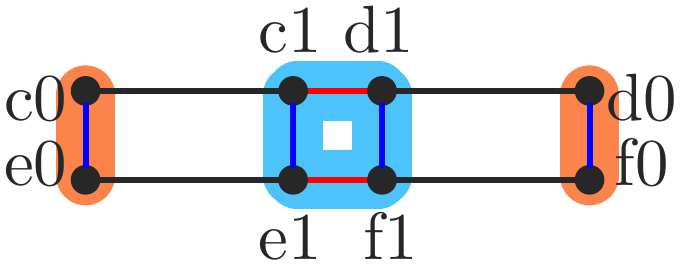}}
    \end{minipage}

    \caption[Instantiating an arc of a graph scheme.]{%
 Instantiating an arc of a graph scheme:
        \protect\subref{ex:instantiationarc:fig:graphscheme} graph scheme with an arc between two nodes,
        \protect\subref{ex:instantiationarc:fig:instantiationfree} instantiation with the orbit graph of \cref{fig:orbit1free}, and
        \protect\subref{ex:instantiationarc:fig:instantiationsewn} instantiation with the orbit graph of \cref{fig:orbit1sewn}.}
    \label{fig:instantiationArc}
\end{figure}

The graph scheme on the orbit \(\orb{\azero,\atwo}\) in \cref{ex:instantiationarc:fig:graphscheme} consists of two nodes linked with a \azero-arc and corresponds to the graph in \cref{fig:discretegraphscheme} with the addition of the \azero-arc.
The instantiation of the nodes has already been discussed.
The next step is the instantiation of the \azero-arc \(n0 \zeroarc n1\), which adds links between the darts corresponding to \(n0\) and \(n1\).
In the case of an initial free edge, node \(n0\) generates two darts, \(a0\) and \(b0\), corresponding to darts \(a\) and \(b\) in the initial orbit graph (see \cref{fig:orbit1free}), while node \(n1\) produces darts \(a1\) and \(b1\). The instantiation adds the links \(a0 \zeroarc a1\) and \(b0 \zeroarc b1\).
We derive a similar construction in the case of an initial sewn edge.
The orbit graph (see \cref{fig:orbit1sewn}) contains four darts, thus nodes \(n0\) and \(n1\) instantiate into four darts each.
The instantiation of \(n0 \zeroarc n1\) creates four links \(a0 \zeroarc a1\), \(b0 \zeroarc b1\), \(c0 \zeroarc c1\) and \(d0 \zeroarc d1\).

\begin{definition}[Arc instantiation -- from~\cite{pascual_inferring_2022}]
    \label{def:arcinstantiation}
 For \(s\) a node of \(\jerboagraphscheme{G}\) and \(u\) a node of \(O\), we write \((u,s)\) for the image of \(u\) in  \(\iota^{\orb{o}}(s, O)\).

 If \(\jerboagraphscheme{G}\) consists of two nodes \(s\) and \(t\) and an arc \(s \larc{i} t\),
 its instantiation with \(O\) extends the instantiation of its nodes to link copies of the same node from~\(O\):
    \[
        \iota^{\orb{o}}(\jerboagraphscheme{G}, O) = \underbrace{\iota^{\orb{o}}(s, O)}_{\text{node }s} ~\cup~ \underbrace{\iota^{\orb{o}}(t, O)}_{\text{node }t} ~\cup~ \underbrace{\bigcup_{u \in O} (u,s) \larc{i} (u,t)}_{\text{arc } s \larc{i} t}
    \]
 We write \(\iota^{\orb{o}}(s \larc{i} t, O)\) for \(\bigcup_{u \in O} (u,s) \larc{i} (u,t)\).

 If \(\jerboagraphscheme{G}\) is a generic graph scheme with node set \(N_{\jerboagraphscheme{G}}\) and arc set \(A_{\jerboagraphscheme{G}}\), its instantiation with \(O\) extends \(\iota^{\orb{o}}(N_{\jerboagraphscheme{G}}, O)\) to link copies according to all the arcs of \(A_{\jerboagraphscheme{G}}\):
    \[
        \iota^{\orb{o}}(\jerboagraphscheme{G}, O) = \underbrace{\iota^{\orb{o}}(N_{\jerboagraphscheme{G}}, O)}_{\text{nodes}} ~\cup~ \underbrace{\bigcup_{(\mu \larc{i} \mu')~\in A_{\jerboagraphscheme{G}}} \iota^{\orb{o}}(\mu \larc{i} \mu',  O)}_{\text{arcs}}
    \]
\end{definition}

To summarize, the instantiation of a graph scheme intuitively corresponds to the following:
\begin{enumerate*}[label=(\arabic*)]
    \item applying each relabeling function from the orbit types decorating the node 
    \item adding a link between the darts image of a node whenever there is an arc in the graph scheme.
\end{enumerate*}

\subsection{Rule Scheme}
The instantiation of a rule scheme \(\jerboarulescheme{L}{\orb{o}}{R}\) is defined as the instantiation of both \(\jerboagraphscheme{L}\) and \(\jerboagraphscheme{R}\) with the same orbit \(O\) of type \(\orb{o}\), resulting in the graph transformation rule:
\[
    \iota^{\orb{o}}(\jerboagraphscheme{L}, O) \to \iota^{\orb{o}}(\jerboagraphscheme{R}, O).
\]

We can finally explain the reconstruction of the two rules of \cref{fig:ajoutSommet1,fig:ajoutSommet3} presented at the beginning of the section.
We consider the rule scheme in \cref{fig:vertexinsertion02folded}. The LHS and RHS are instantiated separately, using the same orbit graph. The instantiation of the RHS has already been discussed.
The LHS consists of a single node \(n0\) with no arcs. Its instantiation is determined by the relabeling function derived from its orbit type, i.e., the identity function \(\orb{\azero, \atwo} \mapsto \orb{\azero, \atwo}\). The instantiations with the graphs in \cref{fig:orbit1free} and~\ref{fig:orbit1sewn} yield exactly these graphs. The right pattern corresponds to the graph scheme in \cref{ex:instantiationarc:fig:graphscheme}.

Therefore, the complete instantiations of the rule scheme correspond to the following:
\begin{itemize}
    \item The instantiation of the left pattern on the graph of \cref{fig:orbit1free} is the graph of \cref{fig:orbit1free}, isomorphic to the LHS of the rule of \cref{fig:ajoutSommet1}.
    \item The instantiation of the left pattern on the graph of \cref{fig:orbit1sewn} is the graph of \cref{fig:orbit1sewn}, isomorphic to the LHS of the rule of \cref{fig:ajoutSommet3}
    \item The instantiation of the right pattern on the graph of \cref{fig:orbit1free} is the graph of \cref{ex:instantiationarc:fig:instantiationfree}, isomorphic to the RHS of the rule of \cref{fig:ajoutSommet1}.
    \item The instantiation of the left pattern on the graph of \cref{fig:orbit1sewn} is the graph of \cref{ex:instantiationarc:fig:instantiationsewn}, isomorphic to the RHS of the rule of \cref{fig:ajoutSommet3}
\end{itemize}

In practice, the orbit type \(\orb{o}\) for the rule parameter is specified by an LHS node called the \emph{hook}. This node, which is required not to have the removing symbol '\(\_\)' in its orbit type, serves as a reference to construct all relabeling functions. Thus, the hook provides the reference for building relabeling functions and indicates where the modeling operation occurs in the object. When applying a rule scheme to a Gmap, the process begins with selecting dart \(a\). From this dart, we build the orbit \(\orb{o}(a)\), where the orbit type \(\orb{o}\) is determined by the hook. The rule scheme is then instantiated with the orbit \(\orb{o}(a)\), creating a graph transformation rule that can be applied to the initial Gmap.

Gmaps correspond to arc-labeled graphs satisfying specific topological constraints, namely the incidence and the cycle constraints. Thus, the instantiation of a graph scheme might never be a subgraph of a Gmap. This situation can be avoided via additional constraints on the graph scheme to ensure that its instantiations are always subgraphs of some Gmap. Having these constraints at the level of graph schemes (and thus rule schemes) minimizes runtime checks in Jerboa when trying to apply an operation defined by a rule scheme. These constraints are beyond the scope of this report but are detailed in~\cite{pascual_topological_2022}.

\section{Conclusion}

Using generalized maps represented as graphs enables graph rewriting mechanisms as a formal foundation for modeling operations. Current approaches to algebraic graph transformation are typically described within a \emph{finitary \(\cM\)-adhesive category} with an \(\cM\)-initial object and \(\cM\)-effective unions (where \(\cM\) is a suitable system of monomorphisms)~\cite{behr_compositionality_2021}, employing compositional double-pushout (DPO) semantics for rewriting~\cite{ehrig_fundamentals_2006}. To circumvent the inherent mathematical complexity of such a framework, this report has presented a set-theoretic explanation for the instantiation of Jerboa rule schemes, offering an alternative perspective to the categorical approach by leveraging the intrinsic proximity between presheaf topoi and sets. The set-theoretic formulation relies on relabeling functions and orbit types to provide a compact yet flexible representation of generic rules. This lightweight formalism serves to bridge the gap between the abstract description of Jerboa rule schemes and their integration with domain-specific requirements.

\printbibliography

\end{document}